\RequirePackage{ifpdf}
\documentclass{JHEP3} 
\pdfoutput=1
\JHEPspecialurl{http://jhep.sissa.it/JOURNAL/JHEP3.tar.gz}
\usepackage{graphicx}
\usepackage{amsmath,amssymb}
\usepackage{cite}
\usepackage{multirow}

\newcommand{\mrm}{\mathrm}
\newcommand{\rd}{\mathrm{d}}
\newcommand{\Br}{\mathcal{B}}
\newcommand{\GeV}{\mathrm{GeV}}
\newcommand{\TeV}{\mathrm{TeV}}
\newcommand{\SM}{\mathrm{SM}}

\newcommand{\alt}{\mathrm{alt}}
\newcommand{\pT}{p_{\mathrm{T}}}
\def\DF1A{\Delta F_{1,A}}
\def\DF1V{\Delta F_{1,V}}
\def\Dphill{\Delta \phi_{ll}}

\def\ttbZ{t\bar{t}Z}
\def\ttb{t\bar{t}}

\def\sw{\sin \theta_w}
\def\swsq{\sin^2 \theta_w}
\def\cw{\cos \theta_w}
\def\invfb {\mathrm{fb}^{-1}}
\def\hLambda {\hat{\Lambda}}
\def\erf{\mathrm{erf}}
\def\ConeA{C_{1,\mathrm{A}}}
\def\ConeV{C_{1,\mathrm{V}}}

\def\ConeVSM{C_\mathrm{V}^{\mrm{SM}}}
\def\ConeASM{C_\mathrm{A}^{\mrm{SM}}}
\def\DConeA{\Delta C_{1,\mathrm{A}}}
\def\DConeV{\Delta C_{1,\mathrm{V}}}
\def\HSM{\mathcal{H}_{\mathrm{SM}}}
\def\Halt{\mathcal{H}_{\mathrm{alt}}}
\newcommand{\be}{\begin{eqnarray}}
\newcommand{\ee}{\end{eqnarray}}

\title{Constraining couplings of top quarks to the $Z$~boson in $t\bar{t}+Z$ production at the LHC}

\author{Raoul R\"ontsch \\ Fermilab, Batavia, IL 60510, USA \\
  Email: \email{rontsch@fnal.gov} }
\author{Markus Schulze \\ PH Department, TH Unit, CERN, 1211 Geneva 23, Switzerland \\
  Email: \email{markus.schulze@cern.ch} }

\received{\today} 		
\accepted{\today}		

\preprint{CERN-PH-TH/2014-053\\
FERMILAB-PUB-14-062-T}

\abstract{
We study top quark pair production in association with a $Z$~boson at the Large Hadron Collider (LHC)
and investigate the prospects of measuring the couplings of top quarks to the $Z$~boson.
To date these couplings have not been constrained in direct measurements. 
Such a determination will be possible for the first time at the LHC. 
Our calculation improves previous coupling studies through the inclusion of next-to-leading order (NLO) QCD corrections in production and decays of all unstable particles.
We treat top quarks in the narrow-width approximation and retain all NLO spin correlations.
To determine the sensitivity of a coupling measurement we perform a binned log-likelihood ratio test
based on normalization and shape information of the angle between the leptons from the $Z$~boson decay.
The obtained limits account for statistical uncertainties as well as leading theoretical systematics from
residual scale dependence and parton distribution functions.
We use current CMS data to place the first direct constraints on the $\ttbZ$ couplings. 
We also consider the upcoming high-energy LHC run and find that with 300~fb$^{-1}$ of data at an energy of 13~TeV the vector and axial $t \bar t Z$ coupling can be
constrained at the 95\% confidence level to $C_\mathrm{V}=0.24^{+0.39}_{-0.85}$ and $C_\mathrm{A}=-0.60^{+0.14}_{-0.18}$, where the central values are the Standard Model predictions.
This is a reduction of uncertainties by 25\% and 42\%, respectively, compared to an analysis based on leading-order predictions.
We also translate these results into limits on dimension-six operators contributing to the $\ttbZ$ interactions beyond the Standard Model.
}

\begin{document}

\section{Introduction}
After run~I of the Large Hadron Collider (LHC) at $\sqrt{s}=7$ and 8~TeV, we look back on a highly successful research program.
Already this first phase of exploring a new energy regime has provided many exciting results: 
the Higgs boson was discovered~\cite{Chatrchyan:2012ufa,Aad:2012tfa}, its quantum numbers and couplings are highly constrained,
and many Standard Model (SM) measurements are competitive with previous ones, if not exceeding them.
Furthermore, a plethora of searches for signals of new physics have been undertaken, reaching out into the multi-TeV region as well as
exploring small deviations of SM parameters. 
The absence of any spectacular signal of new physics highly constrains many minimal extensions of the SM and, at the same time, 
opens up new avenues for
experimental searches and theoretical model building.
These developments represent a remarkably fast progress and demonstrate the potential of the LHC in the years to come.

One particularly promising class of SM processes is top quark pair production in association with gauge bosons or a Higgs boson.
Due to their relatively high production threshold these processes were not accessible at the Tevatron. 
In contrast, the high energy and large luminosity of the upcoming LHC runs will produce sufficiently many events to allow detailed studies of these processes.
Progress in this direction has already been made with cross section measurements of $\ttb+\gamma$ production by ATLAS at 7~TeV~\cite{ATLAS:2011nka}  
and CMS at 8~TeV~\cite{CMS:2014wma}. First events for the processes $\ttb+Z/W$  have also been reported in Refs.~\cite{ATLAS-CONF-2012-126,Chatrchyan:2013qca}.
It is exciting to envision future studies of these processes with direct measurements of the couplings and new sensitivity to physics beyond the Standard Model. 

In this paper we focus on the determination of the top quark to $Z$~boson couplings through $\ttbZ$ production at the LHC. 
This process is a {\it direct} probe of the $\ttbZ$ interactions, which distinguishes it from other indirect probes such as 
the LEP measurements of the $\rho$-parameter~\cite{ALEPH:2005ab} and the $Z \to b \bar{b}$ branching ratio~\cite{Abdallah:2008ab}. 
The SM unambiguously predicts the strength of these couplings, and higher order electroweak corrections modify the leading order values only minimally \cite{Hollik:1988ii}.
On the other hand, extensions of the SM which address, e.g.~dynamic electroweak symmetry-breaking, typically induce larger deviations. 
Popular examples are certain variants of Supersymmetry \cite{PhysRevD.82.055001,PhysRevD.84.015003} or Little Higgs Models~\cite{Schmaltz:2002wx,Cheng:2003ju}.
More generally, any new fermion which mixes with the third generation quarks might induce deviations to the $\ttbZ$ SM couplings. 
Hence also 4th generation quarks~\cite{Frampton:1999xi,Dobrescu:2009vz,Aguilar-Saavedra:2013qpa}, top-color models~\cite{PhysRevD.86.095017,Grojean:2013qca} and extra-dimensional extensions of space-time~\cite{Randall:1999ee,Richard:2013pwa} 
have to be considered.
It is therefore important to know to what extent LHC experiments are sensitive to physics beyond the SM in $\ttbZ$ production.
Clearly, this is not only a question of experimental sensitivity but also depends crucially on our theoretical understanding of the production and decay dynamics 
of the $pp\to\ttbZ$ process.

The ability of LHC experiments to constrain the $\ttbZ$ couplings was first considered in a series of studies by Baur,~Juste,~Orr and Rainwater~\cite{Baur:2004uw,Baur:2005wi}. 
The authors identified suitable observables which are sensitive to vector and axial couplings as well as to the weak electric and magnetic dipole moments.
The tri-lepton signature with semi-hadronically decaying top quarks and a leptonically decaying $Z$~boson turns out to provide a good compromise between
clean signature and large enough cross section. 
But even decay modes with a $Z$~boson decaying into neutrinos yield additional sensitivity~\cite{Baur:2005wi}.
These analyses show that sensitivity to the form factor of the vector current is relatively weak and limits can only be placed within a factor of three with respect to the SM value. 
In contrast, the form factor of the axial current can be pinned down to about 20\% accuracy. 
The authors of Ref.~\cite{Berger:2009hi} perform a similar analysis using the more modern language of effective operators.
This allows them to relate $t\bar{b}W$ and $\ttbZ$ couplings in a combined study of single top and $\ttbZ$ production.

In the context of this work it is important to emphasize that all previous coupling studies were performed at leading-order, and the large residual scale uncertainty 
was identified~\cite{Baur:2004uw} as the main obstacle to stronger constraints on the $\ttbZ$ couplings.
It is the aim of this paper to reduce these uncertainties through a NLO QCD calculation for $\ttbZ$ production and decay into a realistic final state with leptons, jets and missing energy.
The hadronic production of $\ttbZ$, with stable top quarks and a stable $Z$~boson, was previously calculated at NLO QCD accuracy by 
Lazopoulos, McElmurry, Melnikov, and Petriello~\cite{Lazopoulos:2008de}, and  by Kardos, Papadopoulos, and Trocsanyi~\cite{Kardos:2011na}.
The latter calculation was also interfaced to a parton shower~\cite{Garzelli:2011is}, accounting for the decays of the top quarks and $Z$~boson through 
the spin uncorrelated parton shower approximation.
Further hadronization effects were studied in Ref.~\cite{Garzelli:2012bn}.
Since our coupling analysis relies on studying leptonic opening angles we believe that spin correlations are crucial for a correct interpretation of the results.
We therefore account for NLO QCD spin correlations in the decay of top quarks and hadronically decaying $W$~bosons.
This includes the full one-loop corrections as well as soft, collinear and wide angle gluon emission off the top quark decay chain.
Spin correlations of the leptonically decaying $Z$~boson are included as well.
While including all spin correlations, we approximate top quarks and the $Z$~boson as close to on-shell in the narrow-width approximation.
This approximation is parametric in $\Gamma/m$ and its wide range of validity in $\ttb$ production has been studied in Refs.~\cite{Buttar:2008jx,Denner:2012yc,Bevilacqua:2010qb,Heinrich:2013qaa}.

It is interesting to note that the $\ttbZ$ couplings may also be directly probed through single top production in association with a $Z$~boson. 
Indeed, the inclusive cross section of $tZ$ plus its charge conjugate process $\bar{t}Z$ is comparable to the inclusive $\ttbZ$ cross section~\cite{Campbell:2013yla}. 
It turns out that this process is also the leading background to a $\ttbZ$ signal, while other backgrounds such as $pp\to WZ b \bar{b} jj$ are almost negligible~\cite{Baur:2004uw}.
However, it is possible to separate $\ttbZ$ and $tZ$ production by cutting on forward jets and demanding a high jet multiplicity, including two $b$-tagged jets~\cite{Campbell:2013yla}. 
We will therefore consider only the process $pp \to \ttbZ$ in this paper, and defer the study of the couplings using $tZ$ (or a combination of both processes) to a later date.

Finally, let us note that a coupling analysis is not the only scenario in which the process $pp\to\ttbZ$ is interesting. 
The semi-hadronic decay mode of the top quark pairs together with the leptonic $Z$ boson decay is background to several tri-lepton and same-sign lepton searches with additional jets and missing energy.
Such signatures can arise from gluino decays in Supersymmetry, in the context of Universal Extra Dimensions, as well as in models with fermionic top quark partners. 
Furthermore, the invisible decay $Z \to \nu \bar{\nu}$ produces a top pair plus a large amount of missing transverse energy, and is therefore an irreducible background 
to searches for scalar or fermionic top quark partners decaying into top quarks plus dark matter candidates.
While we do not address these topics in this paper, it would be interesting to study the effects of NLO corrections when strong selection cuts are applied on this background.

\section{Outline of the calculation}
In this section, we briefly discuss the features of our calculation.
We consider the tri-lepton signature  
$pp \to \ttb + Z \to t(\to \ell \nu b) \, \bar{t} (\to jj \bar{b}) \, Z(\to \ell \ell)$
which profits from a large cross section due to the hadronic decay of one $W$~boson and the lepton multiplicities from the remaining $W$ and $Z$~bosons.
In our results we will sum over all combination of $e^\pm$ and $\mu^\pm$ in the final state, allowing either $t$ or $\bar t$ to decay leptonically.
Application of the narrow-width approximation for top quarks and the $Z$~boson allows us to separate production and decays stage according to 
\be
  \rd \sigma_{pp\to\ell\ell\ell\nu b \bar{b} jj} = \rd \sigma_{pp\to\ttb+Z} \; \rd\Br_{t\to b \ell\nu} \; \rd\Br_{\bar{t} \to \bar{b} jj} \; \rd\Br_{Z\to \ell\ell}
  \;+\; \mathcal{O}(\Gamma_t/m_t, \, \Gamma_Z/M_Z)
  , \label{Xsec}
\ee
where $\rd \sigma$ denotes the production cross section and $\rd\Br_{X\to Y}= \rd \Gamma_{X\to Y} \big/ \Gamma^\mrm{tot}_X$ are the partial branching fractions.
The use of the narrow width approximation neglects contributions which are parametrically suppressed by $\mathcal{O}(\Gamma / m)$, arising from a largely off-shell top quark or $Z$~boson.
Severe selection cuts on final state particles can violate this approximation when distorting the Breit-Wigner line shape of the resonance.
In our analysis we aim for a large cross section and only place mild cuts required by experimental detector acceptance. 
Hence, we believe the narrow-width is an excellent approximation for our study\footnote{
If necessary we can improve our results by allowing off-shell top quarks, $Z$~boson and photons at LO.
Non-factorizable corrections at NLO QCD which are suppressed by $\alpha_s \, \Gamma/m$ have to be neglected in our framework.
}.
We also neglect the contribution from the decay $t \to Wb+Z$ since the available phase space for on-shell top quarks is tiny and 
$\Br_{t\to W bZ } \approx 3 \times 10^{-6}$~\cite{Altarelli:2000nt,Decker:1992wz,Mahlon:1994us,Jenkins:1996zd}.

\subsection{NLO QCD correction}
At leading order, the production of $\ttbZ$ occurs through the $gg$ and $q\bar{q}$ partonic channels. 
At next-to-leading order QCD, these channels receive real and virtual corrections, while real emission corrections open up the partonic channels $qg$ and $\bar{q}g$. 
We also include NLO QCD corrections to the top quark decays and the hadronically decaying $W$~boson; consequently their total widths are included at LO and NLO as well.
Eq.~(\ref{Xsec}) expanded up to NLO accuracy reads,
\be
  \rd \sigma_{pp\to\ell\ell\ell\nu b \bar{b} jj}^\mrm{NLO} &=& 
  \rd \sigma_{pp\to\ttbZ}^\mrm{LO} \; \rd\Br_{t\to b \ell\nu}^\mrm{LO} \; \rd\Br_{\bar{t} \to \bar{b} jj}^\mrm{LO} \; \rd\Br_{Z\to \ell\ell}
  \; \left( 1 + \chi \right)
  \nonumber \\
  &+&   \rd \sigma_{pp\to\ttbZ+X}^{\delta \mrm{NLO}}  \, \rd\Br_{t\to b \ell\nu}^\mrm{LO} \; \rd\Br_{\bar{t} \to \bar{b} jj}^\mrm{LO} \; \rd\Br_{Z\to \ell\ell}
  \\
  &+&  \rd \sigma_{pp\to\ttbZ}^\mrm{LO} \; \left(  \rd\Br_{t\to b \ell\nu+X}^{\delta\mrm{NLO}} \; \rd\Br_{\bar{t} \to \bar{b} jj}^\mrm{LO} + \rd\Br_{t\to b \ell\nu}^\mrm{LO} \; 
  \rd\Br_{\bar{t} \to \bar{b} jj+X}^{\delta\mrm{NLO}} \right) \rd\Br_{Z\to \ell\ell}
  \nonumber. \label{XsecNLO}
\ee
The factor $\chi= -2 \Gamma_t^{\mrm{tot},\delta\mrm{NLO}}/\Gamma_t^{\mrm{tot,LO}} -2 \Gamma_W^{\mrm{tot},\delta\mrm{NLO}}/\Gamma_W^{\mrm{tot,LO}} $ arises from the $\alpha_s$ expansion
of the total widths in the denominator.
The virtual corrections are evaluated using a numerical OPP realization~\cite{Ossola:2006} of $D$-dimensional generalized unitarity~\cite{Ellis:2007br,Giele:2008ve,Ellis:2008ir} (for a review, see Ref.~\cite{Ellis:2011}).
We extended the framework of Ref.~\cite{Melnikov:2009dn} to account for color neutral bosons,  
which requires new tree level recursion relations as well as an extension of the OPP procedure.
Soft and collinear singularities in the real emission corrections are regularized using the dipole subtraction scheme of Refs.~\cite{Catani:1996vz,Catani:2002hc}, supplemented with a cut-off parameter for the
finite dipole phase space~\cite{Nagy:1998bb,Nagy:2003tz,Bevilacqua:2009zn,Campbell:2010ff}.
The virtual and real corrections to the top quark decay and hadronic $W$ boson decay are implemented analytically. 
Soft and collinear singularities in the real emission decay phase space are regularized using subtraction dipoles given in Ref.~\cite{Melnikov:2011ta}.
We also would like to point out our utilization of parallel computing features. 
We implemented a version of the Vegas integration algorithm which allows parallelization~\cite{pvegas} via the Message-Passing-Interface (MPI)~\cite{mpi-2-standard}. 
The observed speed-up in run time scales almost linearly with the number of CPU cores used. 
This allows us to obtain a full NLO QCD prediction for the total cross section within a few hours on a modern desktop computer with 8 cores.

\begin{figure}[t]
\centering
\includegraphics[width=0.7\textwidth]{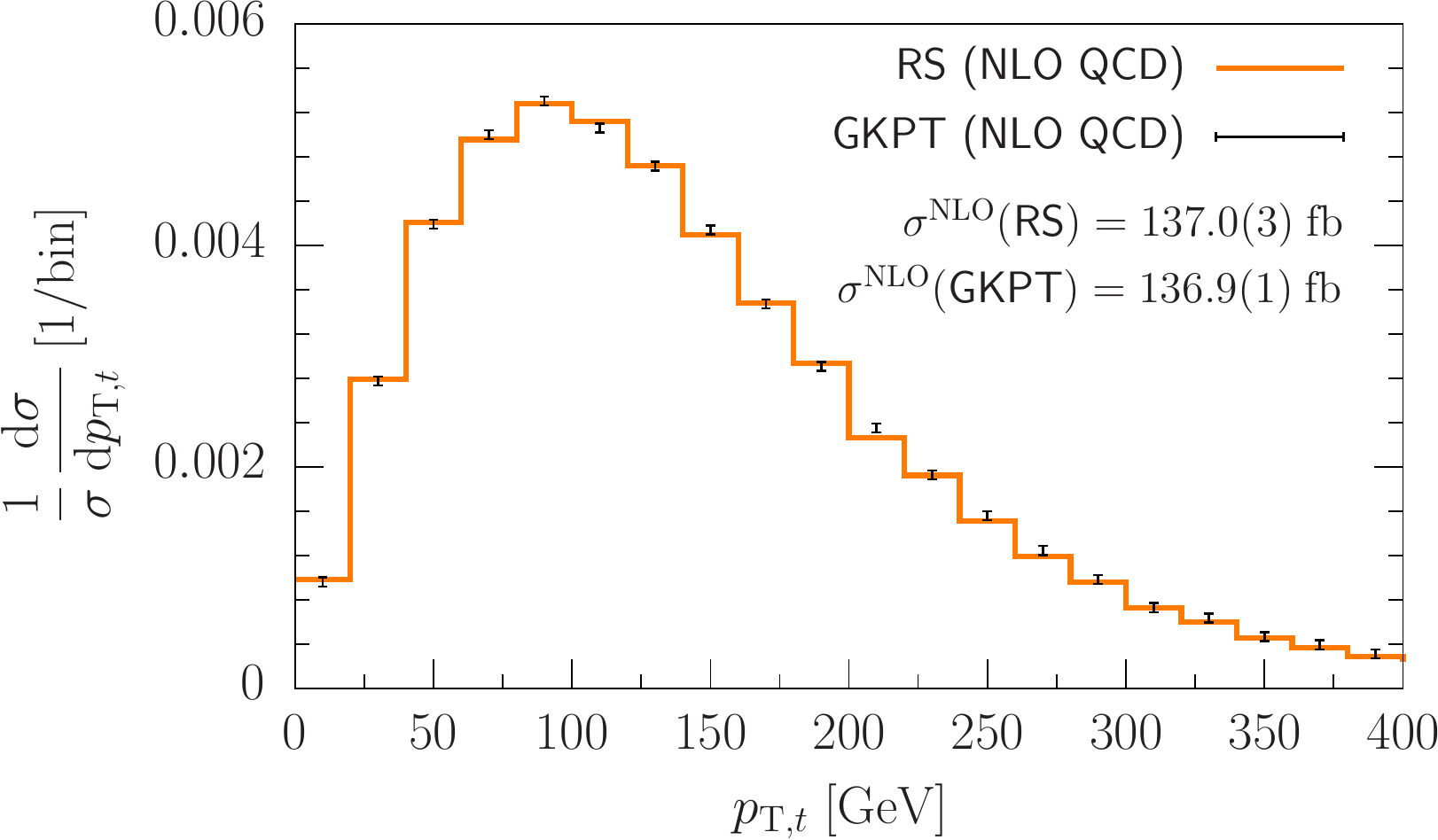}
\caption{\label{fig:i} Shape comparison between our results (RS) and those of Ref.~\cite{Garzelli:2012bn} (GKPT) for 
stable top quarks and $Z$~boson. Shown is the normalized transverse momentum spectrum of the top quark at NLO QCD for the process $pp \to \ttbZ$ at 7~TeV.
}
\end{figure}

We perform several checks to ensure the correctness of our calculation. 
The squared amplitudes for tree level and real emission corrections are checked against {\tt MadGraph v.2.49} ~\cite{Stelzer:1994ta}. 
The cancellation of poles in $D-4$ of dimensional regularization between the virtual corrections and integrated dipoles has been verified for several phase space points.
We also checked the finite part of the virtual amplitudes against the automated program {\tt GoSam}~\cite{Cullen:2011ac} for a few phase space points and find
very good agreement. 
Our framework also allows us 
to turn the $Z$~boson into an on-shell photon which we used to cross check against the amplitudes of Ref.~\cite{Melnikov:2011ta}. 
At the level of the integrated cross section, we vary the cut-off parameter for the finite dipole phase space by at least one order of magnitude and 
verify independence from this parameter for the total cross section and kinematic distributions.
The interface of production and decay amplitudes is checked by integrating over the full phase space and verifying the factorization into the 
inclusive cross section for stable top quarks and $Z$~boson times their branching ratios, at NLO QCD.
Finally, we compare our full hadronic results with a previous calculation~\cite{Garzelli:2012bn} in the literature for stable top quarks and $Z$~boson.
For the purposes of this comparison, we take masses of the top quark, $W$~boson and $Z$ to be $m_t=173.5$ GeV, $M_W=80.39$ GeV, and $M_Z=91.187$ GeV. 
The electroweak coupling is defined through the Fermi constant $G_\mathrm{F}=1.16639 \times 10^{-5} \, \GeV^{-2}$ and the weak mixing angle $\sin^2\theta_w = 1-M_W^2/M_Z^2$. 
CTEQ6L1~\cite{Pumplin:2002vw} and CTEQ6.6M~\cite{Nadolsky:2008zw} parton distribution functions (pdfs) are used at LO and NLO respectively, corresponding to a strong coupling of $\alpha_s^{\mathrm{LO}}(M_Z)=0.130$ and $\alpha_s^{\mathrm{NLO}}(M_Z)=0.118$. 
At the central factorization and renormalization scale of $\mu_0=m_t+m_Z/2$, 
we find a leading order cross section of 103.5(1)~fb and a next-to-leading order QCD cross section of 137.0(3)~fb. 
This is to be compared with the results of Ref.~\cite{Garzelli:2012bn} of 103.5(1)~fb and 136.9(1)~fb, at leading and next-to-leading order QCD.
The cross sections are in excellent agreement within the integration errors.
Figure~\ref{fig:i} also demonstrates good agreement in shape for the top quark $p_{\mathrm{T}}$ distribution between our results and Fig.~1a in Ref.~\cite{Garzelli:2012bn}.

\subsection{$\ttbZ$ couplings}   \label{sect:ttzcoupl}
The $\ttbZ$ interaction Lagrangian in the SM can be written as
\begin{equation} 
  \label{L_SM}
  \mathcal{L}_{\ttbZ}^{\mrm{SM}} = e \, \bar{u}(p_t)\biggl[ \gamma^{\mu} \bigl( \ConeVSM + \gamma_5 \ConeASM \bigr) \biggr]v(p_{\bar{t}}) Z_{\mu},
  \end{equation}
  with the electromagnetic coupling constant $e$. The vector and axial couplings are 
\be
  \ConeVSM &=& \frac{T^3_t - 2Q_t \swsq}{2\sw \cw}, \nonumber \\
  \ConeASM &=& \frac{-T^3_t}{2 \sw \cw},  
\ee
where $Q_t = 2/3$ is the top quark electric charge, $T^3_t=1/2$, and $\theta_w$ is the weak mixing angle. 
The numerical values for the SM couplings are $\ConeVSM \simeq 0.244$ and $\ConeASM \simeq -0.601$.
New physics contributions to the $\ttbZ$ couplings are most conveniently introduced by higher dimensional operators 
in the language of effective field theory. 
A minimal set of dimension-six operators for top quark production and decay have been categorized in Refs.~\cite{AlcarazMaestre:2012vp,AguilarSaavedra:2008zc,Zhang:2012cd}.
In total there are 91 different operators which can be summarized into 20 different anomalous couplings, if on-shellness and gauge invariance is enforced~\cite{AguilarSaavedra:2008zc}. 
For interactions of a $Z$~boson with top quarks only four anomalous couplings, $C_{1/2,V/A}$, remain and Eq.~(\ref{L_SM}) becomes
\be
  \label{L_NP}
  \mathcal{L}_{\ttbZ} = e \bar{u}(p_t)\biggl[ \gamma^{\mu} \bigl(C_{1,V} + \gamma_5 C_{1,A} \bigr)
  + \frac{\mathrm{i} \sigma_{\mu \nu} q_{\nu}}{M_Z} 
  \bigl(C_{2,V} + \mathrm{i} \gamma_5 C_{2,A} \bigr) \biggr] v(p_{\bar{t}}) Z_{\mu} ,
\ee
with $\sigma_{\mu \nu}=\frac{\mathrm{i}}{2} [ \gamma_{\mu},\gamma_{\nu} ]$ and $q_{\nu} = (p_{t}-p_{\bar{t}})_{\nu}$.
In this work we will confine ourselves to the study of the above vector and axial couplings $C_{1,V/A}$, and neglect the $C_{2,V/A}$ terms corresponding to the weak magnetic and electric dipole moments of the top quark. 
Their tree level values vanish in the SM, and $C_{2,V}$ receives one-loop corrections of $ \mathcal{O}(10^{-4})$ \cite{Bernabeu:1995gs}, while $C_{2,A}$ receives finite contributions only beyond two-loops \cite{Hollik:1998vz}. 
On the more technical side, the tensor structure that multiplies the $C_{2,V/A}$ couplings introduces the complication of 
non-renormalizable amplitudes at NLO QCD.
While it is straightforward to handle such contributions, our current implementation of the OPP integrand reduction method 
does not allow tensor ranks larger than $N$ for $N$-point loop integrals.
Such an extension of the OPP reduction algorithm has been outlined in Appendix~B of Ref.~\cite{Mastrolia:2012bu}. 
We intend to come back to this issue in a separate publication in order to study the phenomenological implication of electroweak dipole moments in $\ttbZ$ production.

The couplings $C_{1,V/A}$ can be written in terms of the SM contribution plus deviations due to higher dimensional operators
\be
   \label{Cone_NP}
   &\ConeV=\ConeV^\mathrm{SM}+\frac{1}{4 \sin\theta_w \cos\theta_w}\left(\frac{v^2}{\Lambda^2} \right) \mathrm{Re} \left[ C^{(3,33)}_{\phi q} - C^{(1,33)}_{\phi q} - C^{33}_{\phi u}   \right],
   \\
   &\ConeA=\ConeA^\mathrm{SM}-\frac{1}{4 \sin\theta_w \cos\theta_w}\left(\frac{v^2}{\Lambda^2} \right) \mathrm{Re}\left[  C^{(3,33)}_{\phi q} - C^{(1,33)}_{\phi q} + C^{33}_{\phi u}  \right],
   \nonumber
\ee
where 
\be  
  \label{EFTOp}
  C^{(3,33)}_{\phi q} &=& \mathrm{i} \, (\phi^\dagger \tau^a D_\mu \phi) \, (\bar{t}_\mathrm{L} \gamma^\mu \tau_a t_\mathrm{L})  ,
  \nonumber \\
  C^{(1,33)}_{\phi q} &=& \mathrm{i} \, (\phi^\dagger D_\mu \phi) \, (\bar{t}_\mathrm{L} \gamma^\mu t_\mathrm{L})  ,
 \\
  C^{33}_{\phi u} &=& \mathrm{i} \, (\phi^\dagger D_\mu \phi) \, (\bar{t}_\mathrm{R} \gamma^\mu t_\mathrm{R}).
  \nonumber
\ee
In the above, $t_\mathrm{R,L}$ are the top quark spinors and $\phi$ is the SM Higgs doublet. 
For further definitions, we refer the reader to Ref.~\cite{AguilarSaavedra:2008zc}.
\\

We now would like to comment on existing constraints on the $\ttbZ$ couplings.
Clearly, these constraints are not obtained directly through on-shell production of a $Z$~boson in association with top quark pairs.
Instead, they arise from potential deviations which the higher dimensional operators in Eq.~(\ref{Cone_NP}) introduce to the $\rho$ parameter and the $Z b \bar{b}$ vertex in the SM.
Those parameters are highly constrained through the experimental fits~\cite{Ciuchini:2013pca} of the $\varepsilon$ parameters~\cite{Altarelli:1990zd,Altarelli:1991fk,Altarelli:1993sz},
\be
   \label{epsexp}
   \varepsilon_1^\mathrm{exp} = (5.6 \pm 1.0) \times 10^{-3}, \quad \quad \varepsilon_b^\mathrm{exp} = (-5.8 \pm 1.3) \times 10^{-3}.
\ee
The SM predicts their values as $\varepsilon_1^\mathrm{SM} = (5.21 \pm 0.08) \times 10^{-3} $ and
$\varepsilon_b^\mathrm{SM} = -(6.94 \pm 0.15) \times 10^{-3}$~\cite{Ciuchini:2013pca}.
The new physics contributions in Eq.~(\ref{Cone_NP}) introduce the corrections~\cite{Larios:1999au}
\be
   \delta \varepsilon_1 &=& \frac{3 m_t^2 G_\mathrm{F}}{2\sqrt{2}\pi^2}  
   \, \mathrm{Re}\left[  C^{(3,33)}_{\phi q}-C^{(1,33)}_{\phi q} + C^{33}_{\phi u} + \mathcal{O}\left(\frac{v^2}{\Lambda^2} \right) \right]
   \left( \frac{v^2}{\Lambda^2} \right) \log\left(\frac{\Lambda^2}{m_t^2}\right),
   \\
   \delta \varepsilon_b &=& -\frac{m_t^2 G_\mathrm{F}}{2\sqrt{2}\pi^2} 
   \, \mathrm{Re}\left[  C^{(3,33)}_{\phi q}-C^{(1,33)}_{\phi q} + \frac14 C^{33}_{\phi u}  \right]
   \left( \frac{v^2}{\Lambda^2} \right)\log\left(\frac{\Lambda^2}{m_t^2}\right).
\ee
The experimentally measured values in Eq.~(\ref{epsexp}) can now be used to constrain the operators 
$C^{(3,33)}_{\phi q}$,  $C^{(1,33)}_{\phi q}$ and $C^{33}_{\phi u}$.
We will present the numerical results later in Sect.~\ref{DimSixLimits} together with our results from $\ttbZ$ production.
A further experimental constraint arises from the measurements of the $Z b_\mathrm{L} \bar{b}_\mathrm{L}$ couplings from $R_b$ and $A^{b}_\mathrm{FB}$ at LEP, which are in per-mille level agreement with the SM predictions~\cite{Abdallah:2008ab}.
This experimental fact together with the $\mathrm{SU(2)_L}$ symmetry of the SM
can be used to relate $  C^{(3,33)}_{\phi q} \approx - C^{(1,33)}_{\phi q}$.
Hence, one of the two operators can be eliminated from Eq.~(\ref{Cone_NP}).

\section{Results}
\subsection{NLO QCD Results}
\label{sec:NLOres}
In this section we describe the details of our numerical analysis and the results.
We consider the process 
$pp \to \ttb + Z \to t(\to \ell \nu b) \, \bar{t} (\to jj \bar{b}) \, Z(\to \ell \ell)$
and sum over all combinations of leptons $e^\pm, \mu^\pm$.
We choose the following fixed input parameters
\be
  m_t = 173~\GeV,& \quad   m_b = 0~\GeV,&
  \nonumber\\
  M_Z =91.1876~\GeV,& \quad  M_W =80.385~\GeV,&
  \nonumber\\
  G_\mathrm{F} = 1.166379 \times 10^{-5} \,& \GeV^{-2},  \quad \Gamma_Z = 2.4952~\GeV.&
\ee
Unless otherwise stated, we use MSTW2008 parton distribution functions~\cite{Martin:2009iq} with 
$\alpha_s(M_Z)=0.13939$ and $\alpha_s(M_Z)=0.12018$ at LO and NLO, which we
evolve to the renormalization scale $\mu_\mathrm{ren}$ using 1-loop and 2-loop running, respectively.
The LO and NLO scale dependence has been studied in previous works. 
We do not repeat these studies here, and adopt the central scale $\mu_0=m_t + M_Z/2$ for $\mu=\mu_\mathrm{ren}=\mu_\mathrm{fact}$, as suggested in Ref.~\cite{Lazopoulos:2008de}.
Since we include NLO QCD corrections to the top quark decay and the hadronically decaying $W$~boson, we need to 
include their total widths up to next-to-leading order,
\be
  \Gamma_t^\mathrm{LO} &=& 1.4957~\GeV, \quad \Gamma_t^\mathrm{NLO} = 1.3693~\GeV,
  \nonumber\\
  \Gamma_W^\mathrm{LO} &=& 2.0455~\GeV, \quad \Gamma_W^\mathrm{NLO} = 2.1145~\GeV.
\ee
We consider proton-proton collisions at the LHC with a center-of-mass energy of $\sqrt{s}=13~\TeV$.
To account for detector acceptances and trigger we require
\be
  \label{selectioncuts}
  \pT^{\ell} \ge 15~\GeV, \quad |y^{\ell}| \le 2.5,
  \nonumber\\
  \pT^{j} \ge 20~\GeV, \quad |y^{j}| \le 2.5,
  \nonumber\\
  \pT^{\mathrm{miss}} \ge 20~\GeV, \quad R_{\ell j} \ge 0.4.
\ee
Jet are defined by the anti-$k_\mathrm{T}$ algorithm~\cite{Cacciari:2008gp} with $R=0.4$.
With these input parameters and cuts we find the LO and NLO QCD cross sections,
\be
  \label{XsecNum}
  \sigma_{\ttb Z}^\mathrm{LO} &= 3.79(0)^{+34\%}_{-25\%}~\mathrm{fb},
  \quad\quad\quad
  \sigma_{\ttb Z}^\mathrm{NLO} &= 5.16(1)^{+13\%}_{-12\%}~\mathrm{fb}
\ee
for the central scale $\mu_0$ which is varied by factors of 2 and $1/2$. The value in brackets is the integration error on the last digit.
The dependence on the unphysical scale is reduced from approximately $\pm 30\%$ at LO to $\pm 13\%$ at NLO QCD.
Higher order corrections increase the cross section by 36\%, $K= \sigma_{\ttb Z}^\mathrm{NLO} \big/  \sigma_{\ttb Z}^\mathrm{LO}=1.36$.
\begin{figure}[t]
\centering 
\includegraphics[width=0.49\textwidth]{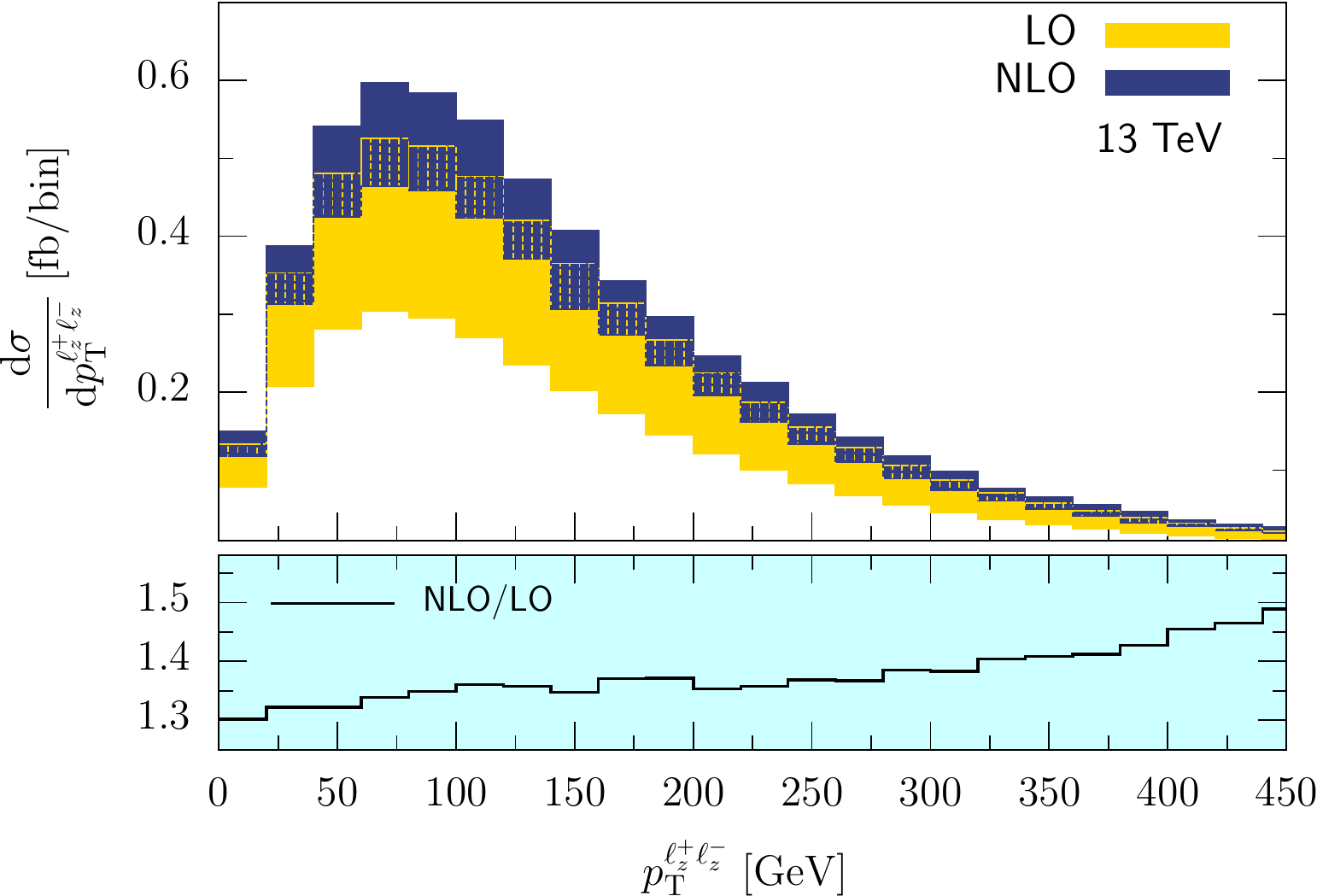}
\includegraphics[width=0.49\textwidth]{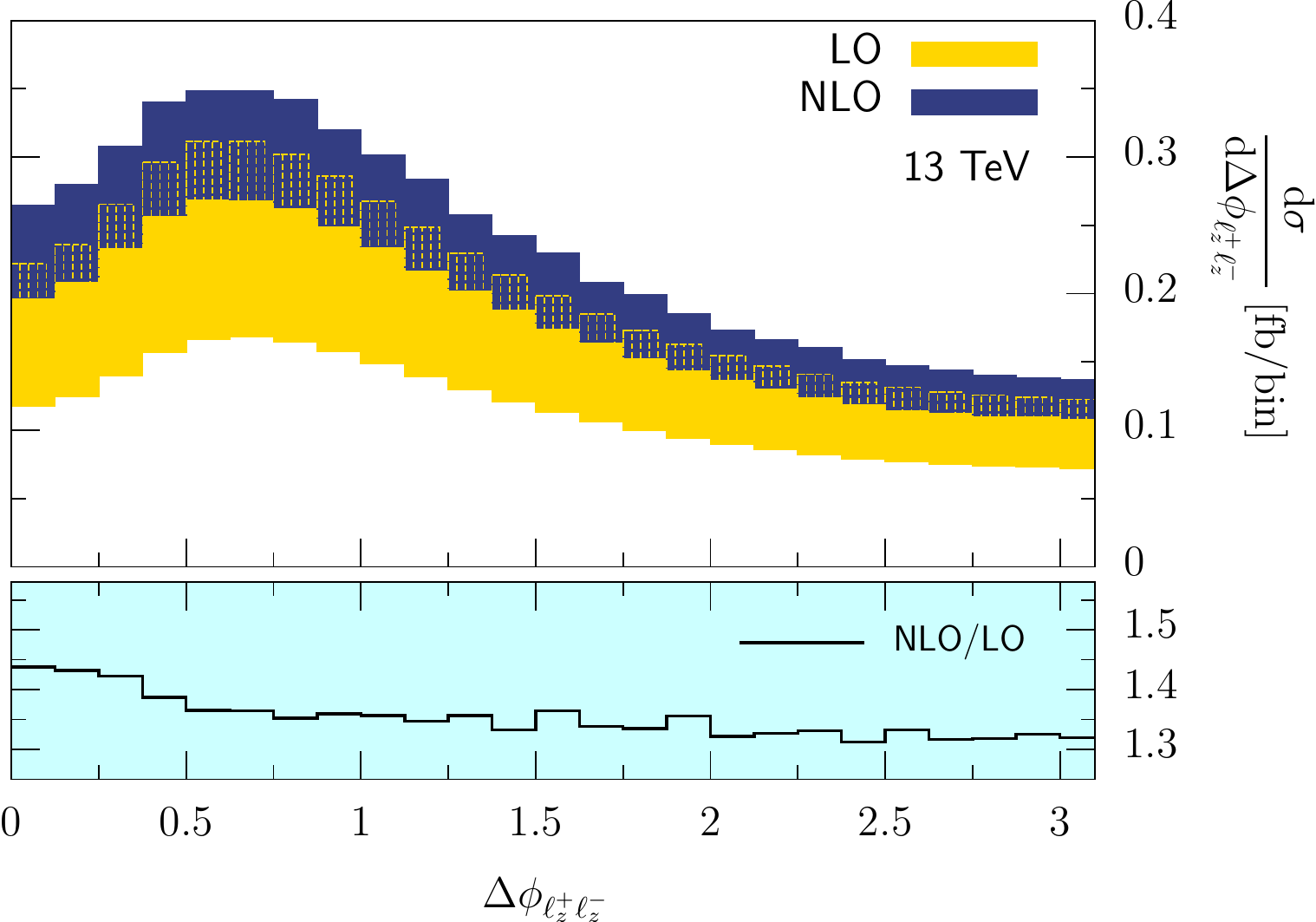}
\caption{\label{fig:ii} Transverse momentum spectrum (left) and azimuthal opening angle (right) of the two leptons from the $Z$ boson 
in the process $pp \to \ttb + Z \to t(\to \ell \nu b) \, \bar{t} (\to jj \bar{b}) \, Z(\to \ell \ell)$ at the 13~TeV LHC.
The bands represent the LO (light) and NLO (dark) results for scale variation by a factor of two around the central scale $\mu_0$.
The lower panes show the differential $K$-factors.
}
\end{figure}
We also calculate the cross sections without any acceptance cuts and
find a significantly lower $K=1.23$.  This emphasizes the importance
of modeling a realistic final state with all unstable particles
decayed.  The ratio of the cross sections with and without cuts
defines the acceptance function $A$, for which we find 
\be
  A^\mathrm{LO} =
  \frac{\sigma_{\mathrm{cuts}}^\mathrm{LO}}{\sigma_{\mathrm{total}}^\mathrm{LO}}
  = 27.1 \%, \quad\quad\quad A^\mathrm{NLO} =
  \frac{\sigma_{\mathrm{cuts}}^\mathrm{NLO}}{\sigma_{\mathrm{total}}^\mathrm{NLO}}
  = 30.0 \%.  
\ee 
The increase of approximately $3\%$ when going from
leading to next-to-leading order seems minor.  However, the common
practice of modeling acceptance effects at LO and multiplying with a
$K$-factor obtained from a NLO calculation with stable particles,
underestimates the correct NLO cross section by $\sim 1-
A^\mathrm{LO}/A^\mathrm{NLO} \simeq 
10\%$.  To estimate uncertainties
from parton distribution functions we contrast the results in
Eq.~(\ref{XsecNum}), using MSTW pdfs, with a calculation that uses the
pdf sets from CTEQ6L1~\cite{Pumplin:2002vw} and CT10~\cite{Lai:2010vv}
at LO and NLO QCD, respectively.  We find \be
\label{XsecNumCTEQ}
  \sigma_{\ttb Z}^\mathrm{LO} &= 3.25(0)^{+34\%}_{-23\%}~\mathrm{fb},
  \quad\quad\quad
  \sigma_{\ttb Z}^\mathrm{NLO} &= 4.80(1)^{+13\%}_{-13\%}~\mathrm{fb}.
\ee
These cross sections are about 14\% smaller at LO and 7\% smaller at NLO QCD compared to the results obtained with MSTW parton distribution functions. 
At NLO, we find that 4\% out of the total difference of 7\% is due to the different values of $\alpha_s(M_Z)$.
The resulting scale uncertainty bands are approximately the same for CTEQ and MSTW pdfs.
Hence, the difference due to the two different parton distribution sets is well within the uncertainty estimate from factorization and renormalization scales.

\begin{figure}[t]
\centering
\includegraphics[width=0.49\textwidth]{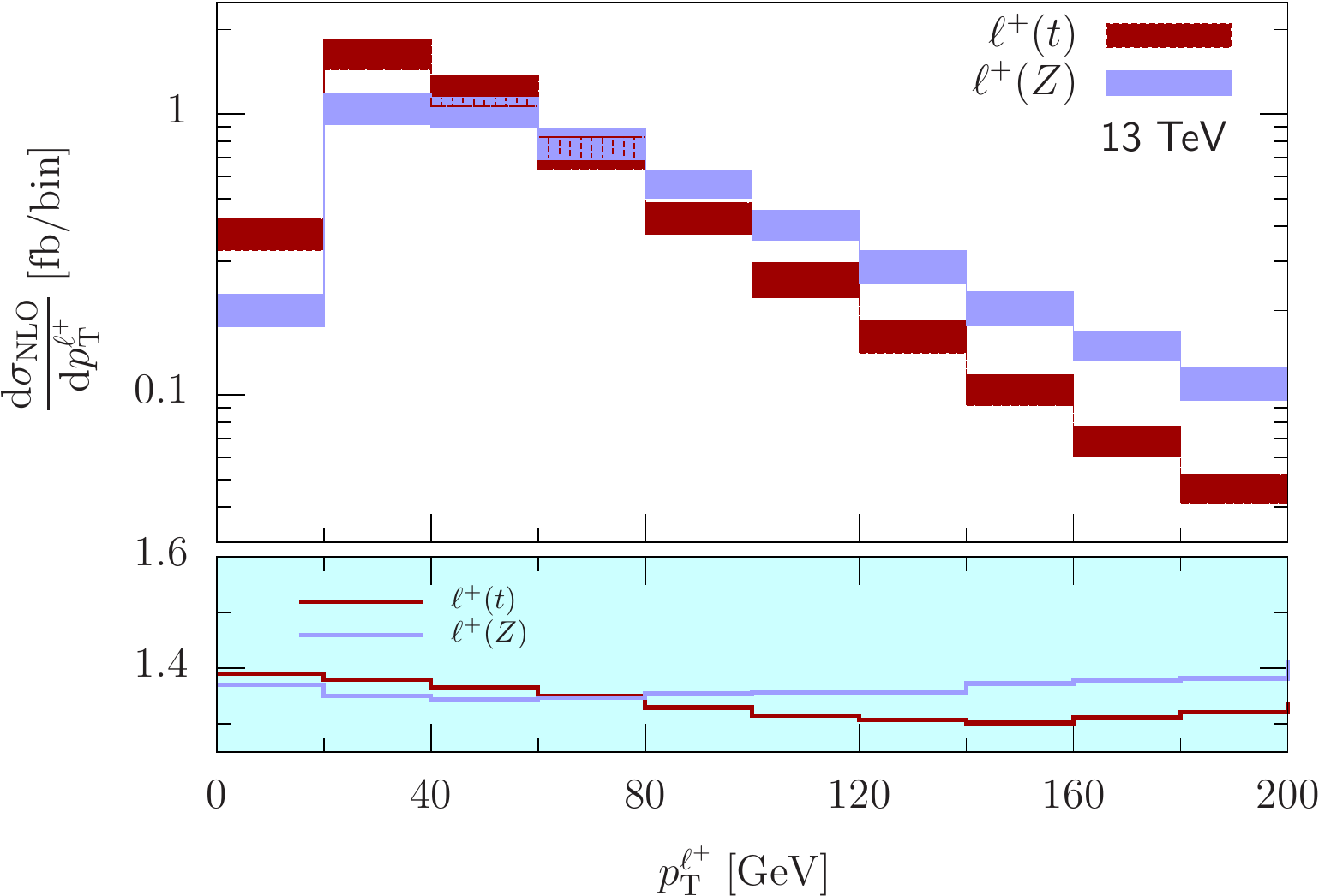}
\hfill
\includegraphics[width=0.49\textwidth]{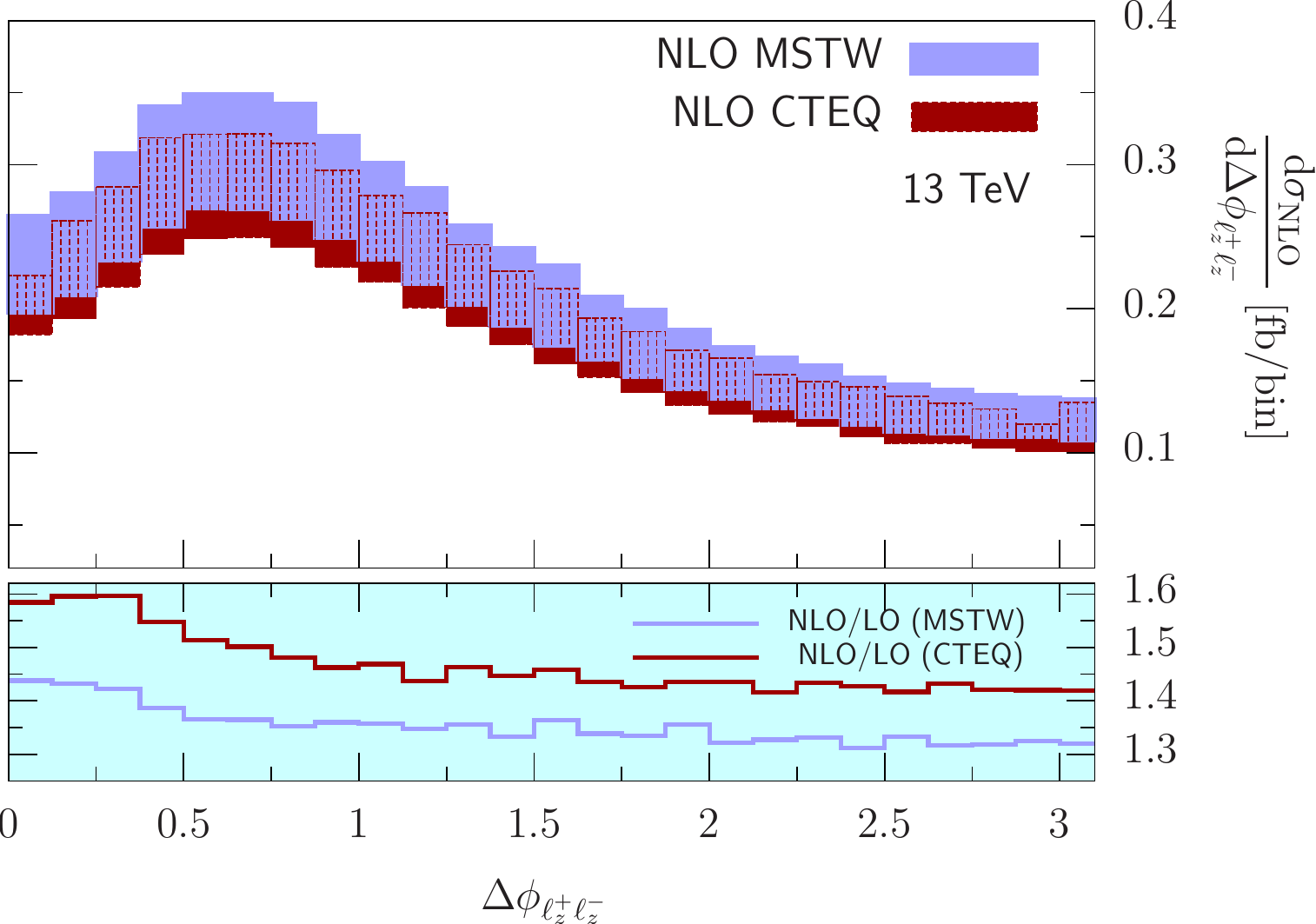}
\caption{\label{fig:iii} Left: Comparison of transverse momenta of leptons arising from the top quark decay (dark) and the $Z$~boson  (light)
in the process $pp \to \ttb + Z \to t(\to \ell \nu b) \, \bar{t} (\to jj \bar{b}) \, Z(\to \ell \ell)$ at NLO QCD.
Right: Comparison of NLO predictions using two different pdf sets (MSTW light, CTEQ dark) for the azimuthal opening angle of the two leptons from the $Z$ boson.
The lower panes show the differential $K$-factors.}
\end{figure}

Before turning to the $\ttbZ$ coupling analysis, let us discuss some generic kinematic distributions.
Fig.~\ref{fig:ii} (left) shows the transverse momentum of the two lepton system reconstructing the $Z$~boson.
Similar to the total cross sections we observe a strong reduction in unphysical scale dependence over the entire $\pT$ spectrum.
Scale bands for LO and NLO predictions are comfortably overlapping. 
From this plot we read off an average transverse momentum of the $Z$~boson of almost 100~GeV with a far-extending kinematic tail,
promising approximately 30 events with $\pT^Z \approx 300~\GeV$ from $300~\mathrm{fb}^{-1}$ at the 13~TeV LHC. 
Fig.~\ref{fig:ii} (right) shows the azimuthal opening angle between the two leptons from the $Z$~boson decay.
This observable has been proven to be a good analyzer of the $\ttbZ$ couplings~\cite{Baur:2004uw} and we will consider it in the following analysis.
Again, we observe strong reduction in scale dependence when going from LO to NLO.
The differential $K$-factor in the lower pane of these plots shows shape changes in the range of 10\% due to higher order corrections, which makes the $\pT^Z$ spectrum harder and decreases the opening angle between the leptons.

On the left hand side of Fig.~\ref{fig:iii} we compare the transverse momentum spectra (at NLO QCD) of the leptons arising from either 
the top quark decay or the $Z$ boson. The leptons arising from the $Z$ boson are significantly harder than those arising from the top decay.
In Fig.~\ref{fig:iii} (right) we study the dependence of our predictions on different parton distribution sets. 
The results for the $\Delta \phi_{\ell^+_z \ell^-_z}$ distribution show that the two NLO predictions 
obtained with MSTW~\cite{Martin:2009iq} and CTEQ~\cite{Pumplin:2002vw,Lai:2010vv} pdfs yield consistent results over the entire spectrum.
However, as can be seen in the lower pane, the $K$-factors differ significantly (10\% or more) 
due to very different predictions with LO pdfs (cf. also Eqs.~(\ref{XsecNum}) and (\ref{XsecNumCTEQ})).

\subsection{Coupling extrapolation and statistical analysis}
\label{sect:analysis}

\begin{figure}[t]
\centering 
\includegraphics[width=0.6\textwidth]{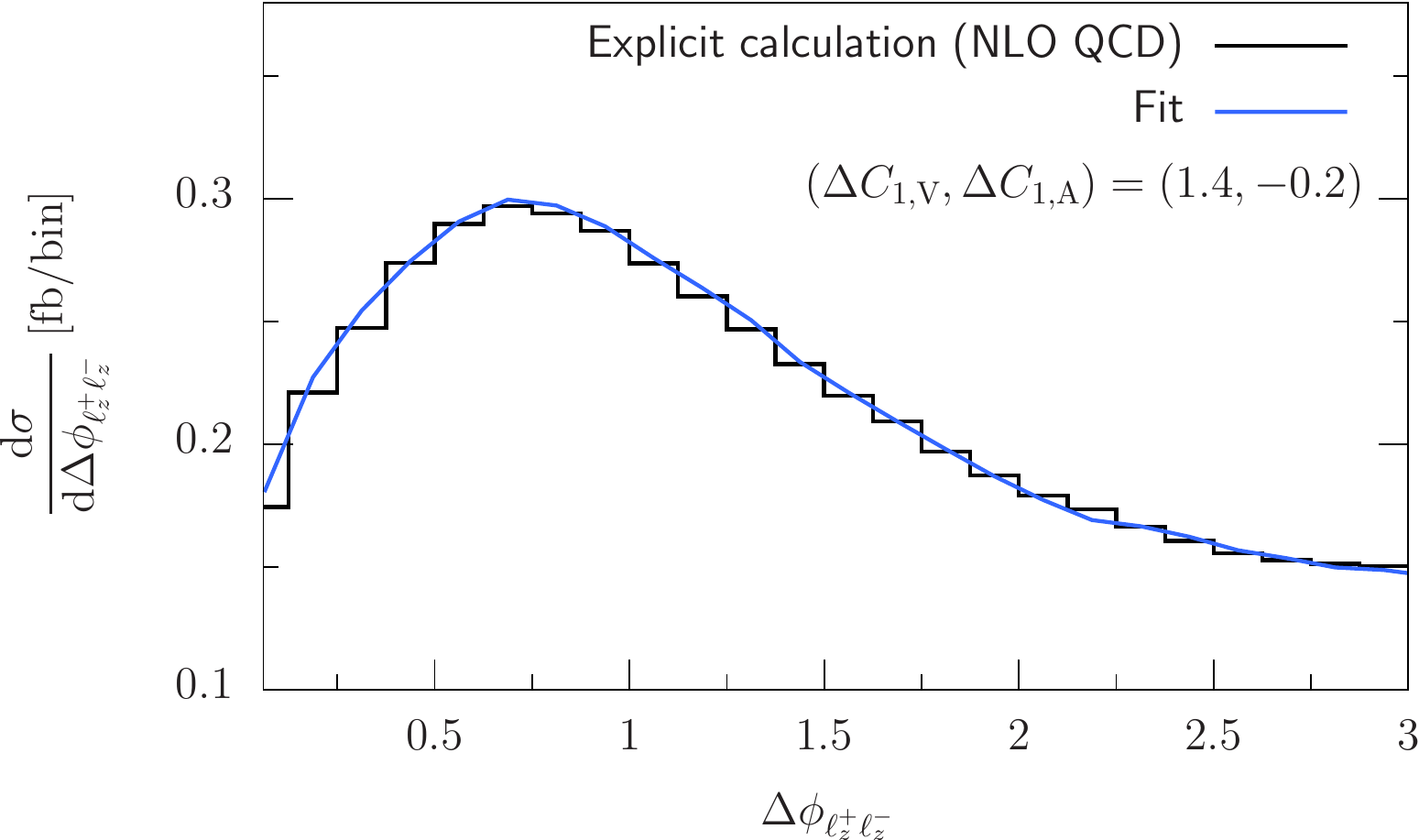}
\caption{\label{fig:iv} Validation of our fitting procedure. Shown is the $\Delta \phi_{\ell^+_z \ell^-_z}$ distribution from an explicit NLO QCD calculation 
for the non-SM coupling choice $(\DConeV,\DConeA)=(1.4,-0.2)$, and from the fit described in Eq.~(3.8). }
\end{figure}

We will now use our calculation to investigate the constraints that can be placed on $\ttbZ$ couplings, 
using both existing and anticipated LHC data. 
To do so, we need to determine how normalization and shapes of differential distributions depend on variations of the couplings. 
Hence total cross sections and differential distributions need to be calculated for a large grid of $C_{1,V}$ and $C_{1,A}$ coupling values. 
This is simple enough at LO, and while it is still feasible at NLO, it does place a strain on computing resources. 
As a convenient alternative, we note that $\ttbZ$ production and decay amplitudes at LO or NLO QCD can be written as
\begin{equation}
    \mathcal{M} = \mathcal{M}_0 +  \ConeV \mathcal{M}_\mathrm{V} +  \ConeA \mathcal{M}_\mathrm{A},
\end{equation}
with the coefficients $\mathcal{M}_i$ encoding both the kinematics and all couplings other than the $\ttbZ$ couplings. 
The differential cross section is then dependent on six coupling structures, and can be written as
\begin{equation}
    \label{couplfit}
    d\sigma = s_0 +s_1C_{1,V} + s_2C_{1,V}^2 +s_3 C_{1,A}+s_4C_{1,A}^2+s_5C_{1,V}C_{1,A}.
\end{equation}
Evaluating the cross section for six values of $(C_{1,V},C_{1,A})$ allows us to solve for the coefficients $s_i$. 
These can then be used to extrapolate results for any values of $C_{1,V}$ and $C_{1,A}$. 
Furthermore, this fitting procedure can not only be done for the total cross section but also bin-by-bin for a given distribution, 
retaining the effects of spin correlations and selection cuts.
As a check of this approach, we have evaluated the cross sections and distributions for a few points in the $(C_{1,V},C_{1,A})$ parameter space, 
both by an explicit calculation and by using the fit for the $s_i$ coefficients. Excellent agreement is found in all cases. 
As an example, we show one comparison in Fig.~\ref{fig:iv} for the  $\Delta \phi_{\ell^+_z \ell^-_z}$ distribution, which we will later use in the coupling analysis.
As can be seen in Fig.~\ref{fig:iv}, the overall normalization and the shape are correctly reproduced by the fitting procedure.
In this figure and the following, the relative shifts in the couplings are given by
\be
  \DConeV =  \frac{\ConeV}{\ConeVSM}-1,
  \hspace{2cm} 
  \DConeA = \frac{\ConeA}{\ConeASM}-1.
\ee

\begin{figure}[t]
\centering 
\includegraphics[width=0.49\textwidth]{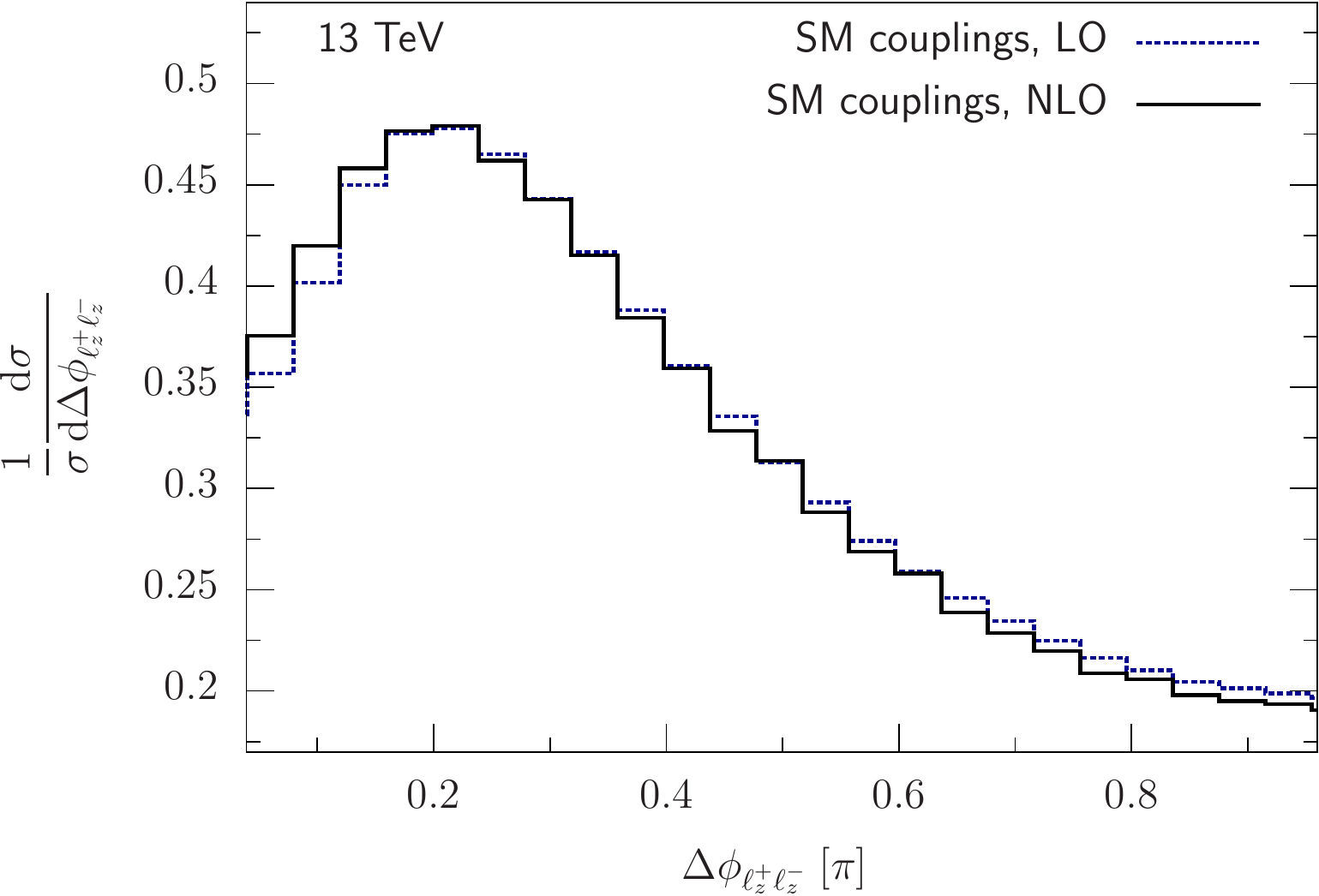}
\hfill
\includegraphics[width=0.49\textwidth]{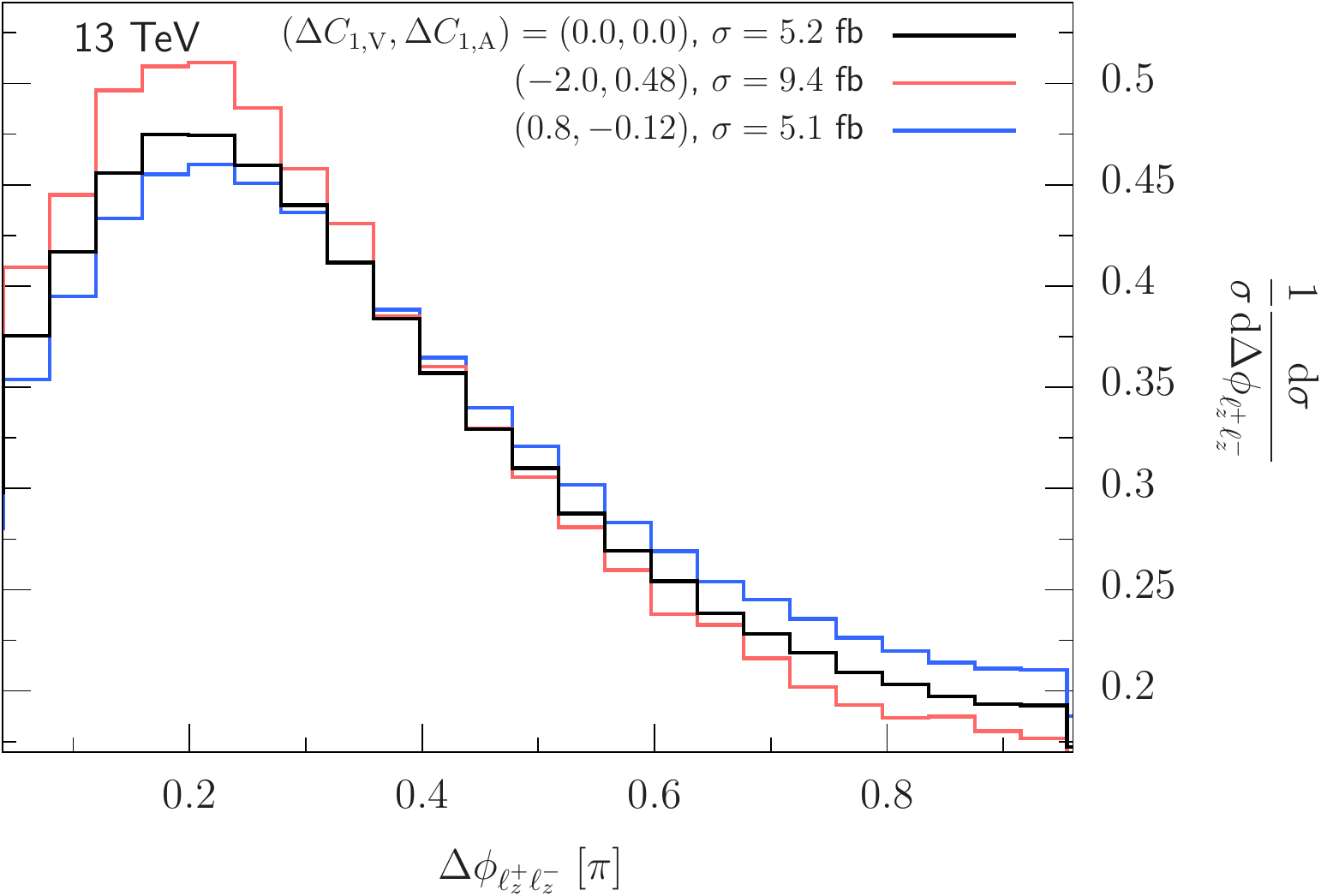}
\caption{\label{fig:vi}
Normalized distributions of the azimuthal opening angle of the opposite sign leptons from the $Z$~boson decay at the 13~TeV LHC.
In the left plot, shapes of LO and NLO QCD predictions are compared for SM $\ttbZ$ couplings.
Shape changes due to deviations from the SM values are shown in the right plot, for the NLO calculation.}
\end{figure}

In our analysis we focus on the tri-leptonic final state and employ the azimuthal angle between the leptons originating
from the $Z$ boson decay to perform our analysis.
This angle has been identified as being particularly sensitive to the $\ttbZ$ couplings in Ref.~\cite{Baur:2004uw}.
We have already discussed the strong reduction in scale uncertainty when going from LO to NLO QCD for this observable.
Here, in Fig.~\ref{fig:vi} (left), we show the effect of NLO QCD corrections on the shape of the normalized $\Delta \phi_{\ell\ell}$ distribution.
Higher order effects tend to shift events from larger to smaller opening angles.
In Fig.~\ref{fig:vi} (right) we show that similar shape changes can arise due to variations of the vector and axial $\ttbZ$ couplings.
This emphasizes the importance of precise predictions, since missing higher order effects might be misinterpreted as deviations from the SM.
To illustrate that the $\Delta \phi_{\ell\ell}$ shape is a useful discriminator for our coupling analysis, we have chosen 
a value $(\DConeV,\DConeA)$ in Fig.~\ref{fig:vi} (right) such that the total cross section approximately coincides with the SM $\ttbZ$ cross section.
Hence, a measurement of the rate alone would not reveal the deviations from its Standard Model value.

Let us now outline the basic features of our statistical analysis.
We are interested in answering the question: what are the bounds that can be placed on deviations of the $\ttbZ$ couplings, assuming that the SM is true?
Obviously, the answer will depend on the assumed integrated luminosity of the data sample as well as on theoretical and experimental uncertainties. 
We assume the SM prediction as our null hypothesis 
$\mathcal{H}_{\mathrm{SM}}$ ($\DConeV,\DConeA)=(0,0)$,
against which we test alternative hypotheses $\Halt$ with $(\DConeV,\DConeA) \ne (0,0)$.
Alternatively, the null hypothesis can be replaced by observed data from LHC experiments to determine the best fit in $(\DConeV,\DConeA)$ parameter space.
This also enables the exclusion of parts of the parameter space at a given confidence level. 
Assuming that the data agree with the SM, 
the bounds on the $\ttbZ$ couplings 
obtained in this work should approximate those obtained from real data. 
We construct two likelihood functions $\mathcal{L}_{\SM}$ and $\mathcal{L}_{\alt}$ which allow us to define a test 
statistic $\Lambda = \log \left( \mathcal{L}_{\SM} / \mathcal{L}_{\alt} \right)$.
We then generate two event samples for a fixed integrated luminosity assuming that either $\mathcal{H}_{\mathrm{SM}}$ or $\Halt$ is true.
The test statistic $\Lambda$ can be evaluated for these two event samples, and 
repeating this evaluation in a large number of pseudo experiments provides the probability distributions $P(\Lambda|\mathcal{H}_{\mathrm{SM}})$ and $P(\Lambda|\Halt)$.
The overlap of these two probability distributions can be used to define the type-I error for rejecting $\mathcal{H}_{\mathrm{SM}}$ in favor of $\Halt$, even though $\mathcal{H}_{\mathrm{SM}}$ is true.
This error can finally be translated into the more familiar confidence level in terms of standard deviations.

Let us now describe the procedure outlined above more precisely and illustrate how differential distributions at NLO QCD can be used.
We closely follow typical likelihood-based analyses as described for example in Ref.~\cite{Cowan:2010js}, based on the original procedure by 
Feldman and Cousins~\cite{Feldman:1997qc}.
The starting point is the binned likelihood function 
\be
   \label{lili}
   \mathcal{L}(\mathcal{H}|\vec{n}) = \prod_{i=1}^{N_\mathrm{bins}} P_i(n_i|\nu_{i}^\mathcal{H})
\ee
with the Poisson distribution $P_i$ for $n_i$ events in the $i$-th bin, given the expected value $\nu_{i}^\mathcal{H}$ for hypothesis $\mathcal{H}$. 
Consequently the two log-likelihood functions for the SM and the alternative hypothesis read
\be
  \label{lilifunct}
  \log\mathcal{L}(\HSM |\vec{n}_\mathrm{obs})  &=& \sum_{i=1}^{N_\mathrm{bins}} \bigl[ n_{i,\mathrm{obs}}\log(\nu_i^{\SM}) -\log(n_{i,\mathrm{obs}}!) -\nu_i^{\SM}  \bigr], 
  \nonumber\\
  \log\mathcal{L}(\Halt|\vec{n}_\mathrm{obs})  &=& \sum_{i=1}^{N_\mathrm{bins}} \bigl[ n_{i,\mathrm{obs}}\log(\nu_i^{\alt})-\log(n_{i,\mathrm{obs}}!) -\nu_i^{\alt} \bigr],
\ee
where the sum over $i$ runs over all bins in a given histogram.
Eqs.~(\ref{lilifunct}) allow us to construct a log-likelihood ratio which serves as the test statistic
\be
  \Lambda(\vec{n}_\mathrm{obs}) &=& \log \biggl( \mathcal{L}(\HSM |\vec{n}_\mathrm{obs})  \big/ \mathcal{L}(\Halt|\vec{n}_\mathrm{obs})  \biggr)  
  \nonumber \\
                                &=& \sum_{i=1}^{N_\mathrm{bins}} \biggl[ n_{i,\mathrm{obs}}\log \biggl( \frac{\nu_i^{\SM}}{\nu_i^{\alt}} \biggr) -\nu_i^{\SM} + \nu_i^{\alt} \biggr].
\ee
The log-likelihood ratio is guaranteed to be the optimal test statistic through the Neyman-Pearson lemma ~\cite{NeymanPearsonLemma}. 
It can now be evaluated with $\nu_i$ being the events from the $\Dphill$ histogram, and $\vec{n}_{\mrm{obs}}$ being the Poisson distributed events around $\nu_i$, for either the SM or the alternative hypothesis.
\begin{figure}[t]
\centering 
\includegraphics[width=0.49\textwidth]{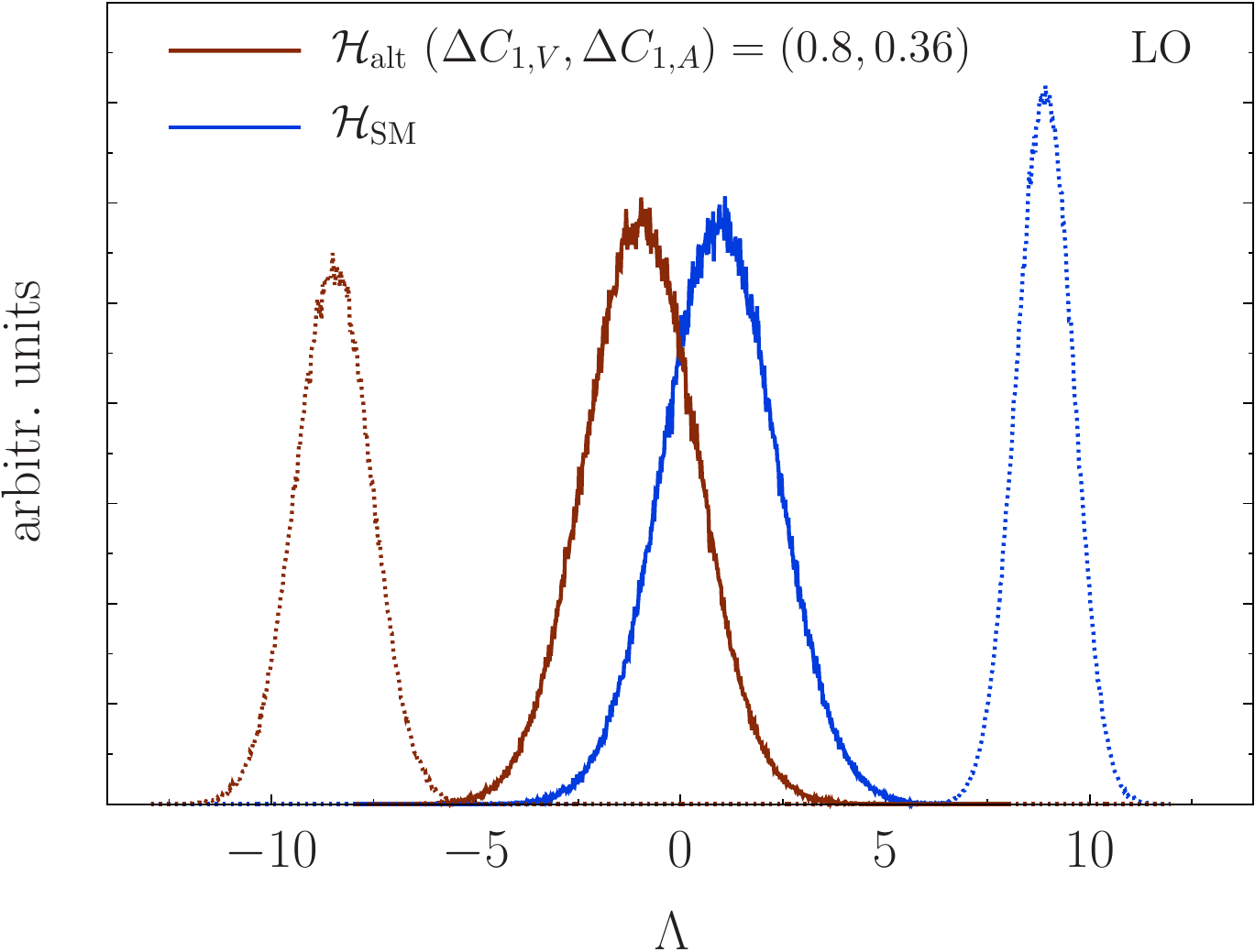}
\hfill
\includegraphics[width=0.49\textwidth]{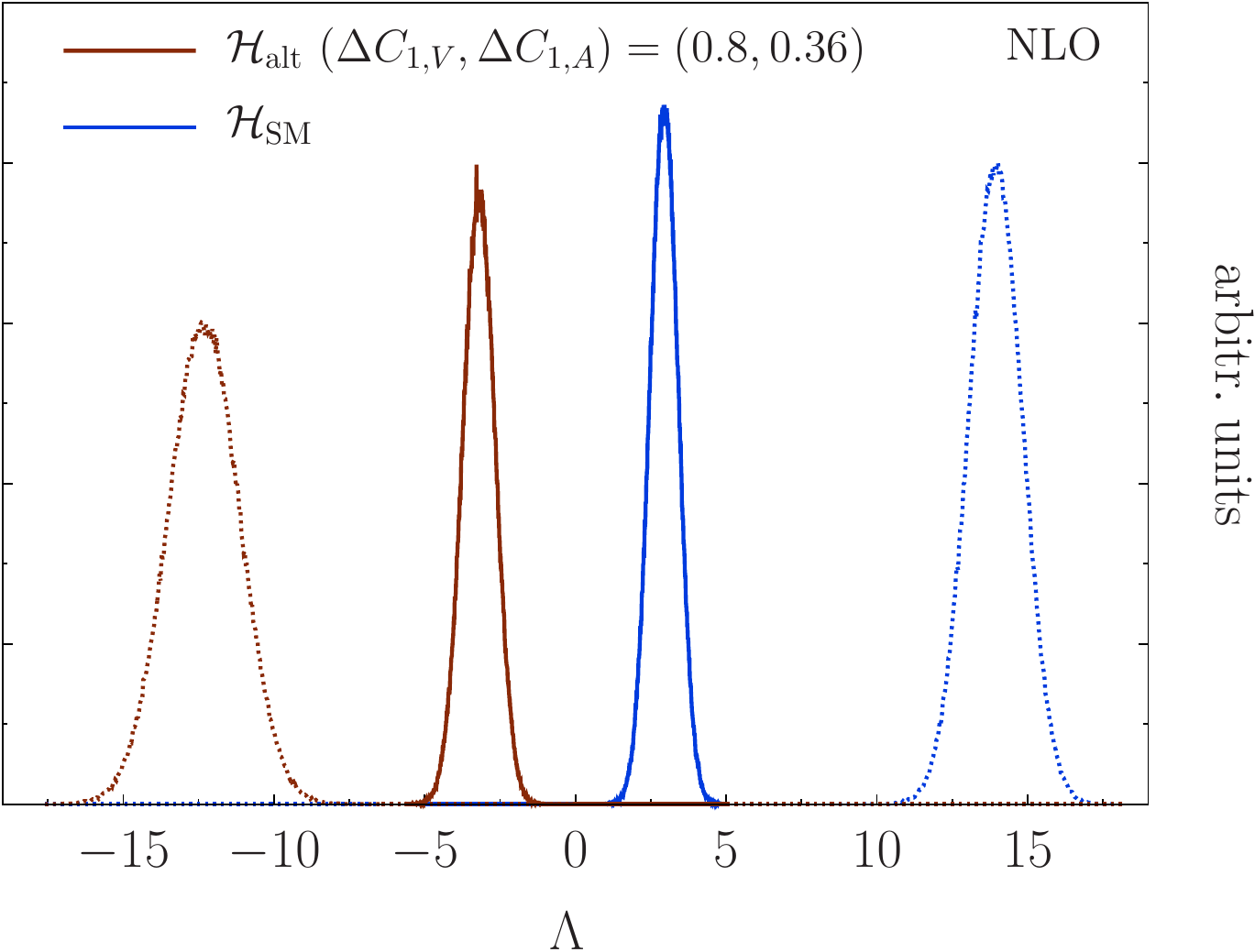}
\caption{\label{fig:vii}
Probability distributions of the log-likelihood ratio $\Lambda$ assuming that the observed events follow the SM hypothesis (blue) or an alternative hypothesis (red) 
with $(\DConeV,\DConeA)=(0.8,0.36)$.
The solid lines include statistical and systematic uncertainties as described in the text, whereas the dashed lines only include statistical uncertainties.
The left plot shows the separating power using LO input with $\Delta_\mathrm{syst.}=30\%$, 
the right plot is obtained at NLO QCD with $\Delta_\mathrm{syst.}=15\%$, assuming $\sqrt{s}=13\,\TeV$ and $\mathcal{L}=300\,\invfb$.
}
\end{figure}
Repeating this procedure for a large number of pseudo-experiments yields the two probability distributions of $\Lambda(\vec{n}_\mathrm{SM})$ and $\Lambda(\vec{n}_\mathrm{alt})$. 
An example of two such probability distributions, $P(\Lambda|\mathcal{H}_{\mathrm{SM/alt}})$, is shown in Fig.~\ref{fig:vii} for LO~(left) and NLO QCD~(right).
These two distributions can be used to define a confidence level for excluding the alternative hypothesis.
For a given value $\hat{\Lambda}$, the probability of accepting $\mathcal{H}_{\alt}$ even though $\mathcal{H}_{\SM}$ is correct (type-I error) is
\begin{equation}
    \alpha = \int^{\hLambda}_{-\infty} \mathrm{d}\Lambda \; P(\Lambda | {\HSM}).
\end{equation}
Similarly, the probability of accepting $\mathcal{H}_{\SM}$ even though $\mathcal{H}_{\alt}$ is correct (type-II error) is given as 
\begin{equation}
    \beta = \int_{\hLambda}^{\infty} \mathrm{d}\Lambda \; P(\Lambda|\Halt).
\end{equation}
We define $\hLambda$ such that $\alpha=\beta$, i.e. there is equal chance of {\it incorrectly} rejecting one hypothesis in favor of the other. 
The value $\alpha(\hLambda)$ is then a measure of statistical discrimination between the two hypotheses. 
It can be translated into the more familiar number of standard deviations by 
\begin{equation} \label{alphatosig}
    \sigma = \sqrt{2} \, \erf^{-1}(1-\alpha),
\end{equation}
where $\erf^{-1}$ is the inverse error function. 

The above discussion made no mention of systematic uncertainties.
In this work we would like to include the leading theoretical uncertainties from unphysical scale dependence and errors associated with parton distribution functions.
For simplicity we neglect experimental systematics such as efficiencies or momentum smearing effects.
Note however that we include realistic detector acceptances through the cuts in Eq.~(\ref{selectioncuts}).
Statistical fluctuations are obviously included in our analysis through the Poisson distribution in Eq.~(\ref{lili}).
Following Ref.~\cite{Conway:2011in}, we include the theoretical uncertainties through nuisance parameters by including multiplicative factors in the log-likelihood function.
The inclusion of nuisance parameters removes the Neyman-Pearson guarantee that the log-likelihood ratio is the optimal test statistic.
Nevertheless, one still expects the test to be approximately optimal as long as the nuisance parameters are reasonably constrained.
We include the theoretical uncertainties by modifying the likelihood function in Eq.~(\ref{lili}) according to
\be
  \label{errorfunctG}
  \mathcal{L}(\mathcal{H}|\vec{n}) \to \mathcal{L}(\mathcal{H}|\vec{n})  \,\times\, \mathcal{G} \left( \nu_i^\mathcal{H} | \tilde{\nu}_i^\mathcal{H}(\Delta_\mathrm{theor.\,unc.}) \right).
\ee
We choose $\mathcal{G}$ to be a constant normalized function with support $\tilde{\nu}_{\mathrm{min/max},i}^\mathcal{H} = \tilde{\nu}_i^\mathcal{H}(1 \pm \Delta_\mathrm{theor.\,unc.}$),
\be
\label{uniuncert}
  \mathcal{G} \left( \nu_i^\mathcal{H} | \tilde{\nu}_i^\mathcal{H}(\Delta_\mathrm{theor.\,unc.}) \right) = 
  \big( \theta\left( {\nu}_i^\mathcal{H} - \nu_{\mathrm{min},i}^\mathcal{H}  \right) 
      \times  \theta\left( \nu_{\mathrm{max},i}^\mathcal{H} - {\nu}_i^\mathcal{H} \right)  \big) 
  \big/ \big( \nu_{\mathrm{max},i}^\mathcal{H} - \nu_{\mathrm{min},i}^\mathcal{H}  \big),
\ee
where $\theta(x)$ is the unit step function.
The value of $\tilde{\nu}_i^\mathcal{H}$ is determined by the assumed luminosity times the cross section in the $i$th bin 
for the central scale choice $\mu_0$.
To be most conservative, we choose the cross section within the uncertainty band for each hypothesis such that their difference is minimized. 
The value in each bin is then rescaled accordingly.
This treatment results in a larger overlap between the two likelihood distributions, 
and consequently a larger $\alpha$ value and less discriminatory power between the two hypotheses. 
This feature is clearly visible in Fig.~\ref{fig:vii}, when comparing the solid with the dotted curves. 
Contrasting the LO results in Fig.~\ref{fig:vii} (left) with the NLO result (right) shows that 
the lower uncertainty associated with the NLO prediction allows for significantly better statistical discrimination between the hypotheses.
Additionally, the increase of the NLO cross section due to the $K$-factor of approximately 1.4 leads to a larger number of expected events and therefore 
to smaller statistical uncertainties.

\subsection{$\ttbZ$ coupling constraints from current and future LHC data}
\label{sect:CouplLimits}

\begin{figure}[t]
\centering
\includegraphics[scale=0.6]{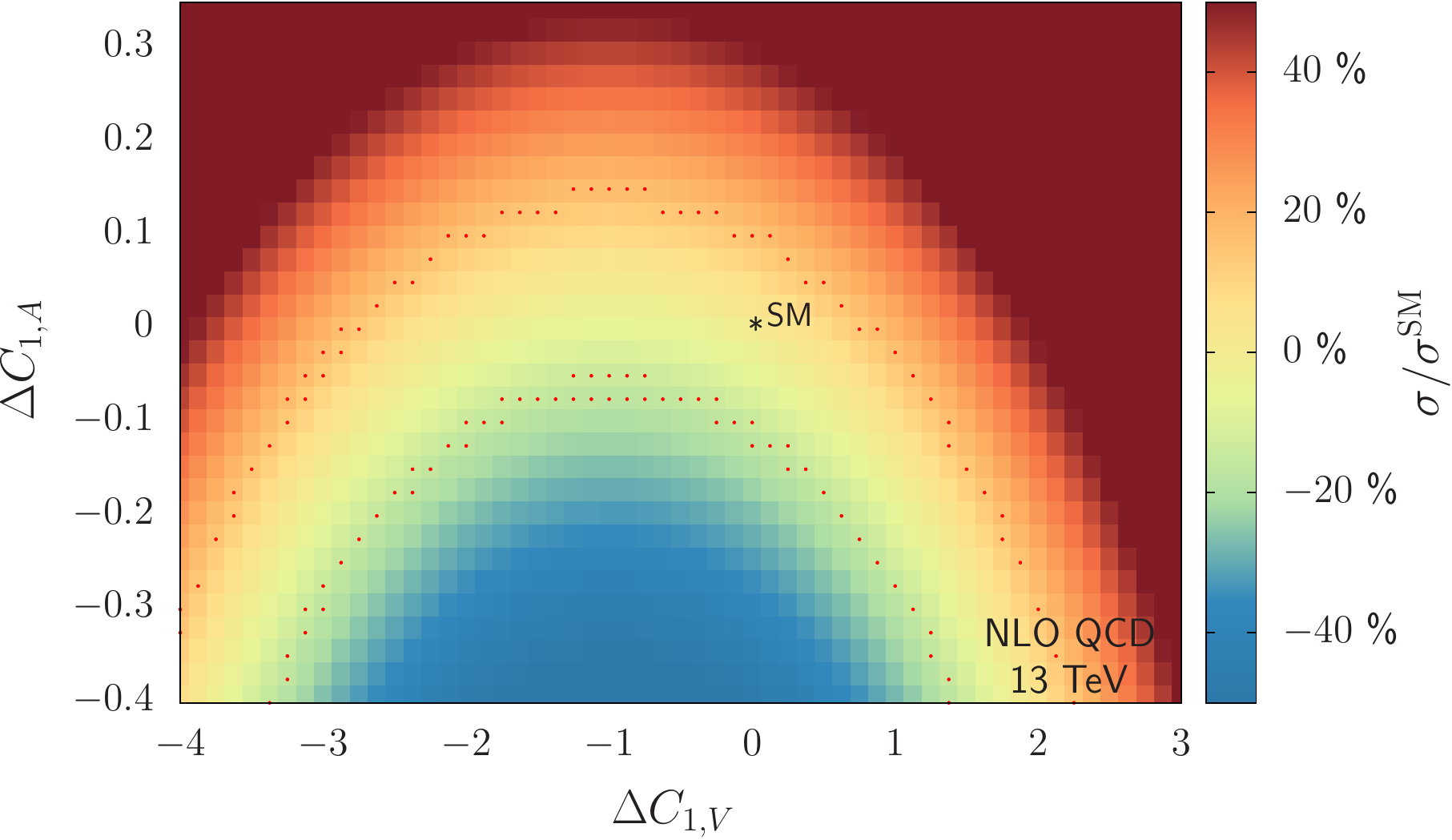}
\caption{ \label{fig:viii} Relative deviations of the NLO QCD cross section as a function of relative shifts in vector and axial couplings with respect to the SM.
The grid of $ 80 \times 40 $ coupling choices is obtained from the fit described in Eq.~(3.8). }
\end{figure}

We now apply the analysis outlined in the previous section to study coupling constraints from current and future LHC data.
Figure~\ref{fig:viii} shows the relative shift of the $\ttbZ$ cross section for non-SM couplings with respect to the SM cross section 
for a wide range of vector and axial couplings. 
The grid of 3200 NLO QCD cross sections is generated with the fit described in Eq.~(\ref{couplfit}), at low computational cost, 
and accounts for selection cuts of Eq.~(\ref{selectioncuts}).
We find that within the given range the cross sections vary by about $\pm 50\%$ away from the SM value due to shifts of vector and axial couplings.
The remaining scale uncertainty at NLO QCD was found to be $\pm 13\%$, which roughly corresponds to the area enclosed by the dotted line in Fig.~\ref{fig:viii}.
Hence, for all coupling values within this band, a rate measurement alone is not sensitive to any deviation.
This is true for a large range of couplings far off the SM value, e.g. $(\Delta\ConeV,\Delta\ConeA)=(1.7,-0.3)$.
We will later see that adding shape information from kinematic distributions will improve this situation and lead to a more powerful discrimination. 
It is clearly noticeable in Fig.~\ref{fig:viii} that cross sections are symmetric around the axis $\Delta\ConeV=-1$. 
This feature can be easily understood from the fact that the LO cross section is dominantly proportional to 
$\ConeV^2+\ConeA^2$ and $\Delta\ConeV=-1$ corresponds to the point $\ConeV=0$.
We expect to see a similar symmetry around $\Delta\ConeA=-1$, however  the sign of the axial coupling is already constrained from 
LEP measurements of the $Zb_\mathrm{L}\bar{b}_\mathrm{L}$ interaction when $\mathrm{SU(2)}_\mathrm{L}$ symmetry is invoked, and consequently we do not show the results for 
a negative value of $\ConeA$ here.

As a side remark we would like to briefly comment on the effects of $\ttbZ$ coupling shifts on the top quark 
forward-backward asymmetry ($A_\mathrm{FB}^{\ttb}$) at the Tevatron.
It is known that the QCD induced asymmetry of $A_\mathrm{FB}^{\ttb}\approx 5\%$ \cite{Kuhn:2011ri}
is enhanced by $+1$\% through the parity-violating electroweak process $q \bar{q} \to Z/\gamma^* \to \ttb$ \cite{Kuhn:2011ri,Hollik:2011ps}.
This enhancement could potentially be larger due to large deviations of the $\ttbZ$ couplings from their SM values, as assumed in this paper.
We checked that for coupling variations within the range shown in Fig.~\ref{fig:viii},
the electroweak contribution to $A_\mathrm{FB}^{\ttb}$ is not significantly altered. 
Hence, any discrepancy between theory and experiment for $A_\mathrm{FB}^{\ttb}$ at the Tevatron cannot be explained by deviations of the $\ttbZ$ couplings as
assumed in this paper.

We now use current LHC data to obtain direct constraints on vector and axial couplings. The production of $\ttbZ$ has been observed at the $\sqrt{s}=7$ TeV run at the LHC, 
with CMS observing nine events~\cite{Chatrchyan:2013qca}, and ATLAS observing one event with more stringent selection criteria~\cite{ATLAS-CONF-2012-126}. 
This enables ATLAS to place an upper bound on the $\ttbZ$ cross section, 
while CMS is able to determine $\sigma_{\ttbZ}^{\mathrm{CMS}} = 0.28^{+0.14}_{-0.11}\mathrm{(stat.)}~^{+0.06}_{-0.03}\mathrm{(sys.)}$~pb. 
Clearly, error bars from this very first measurement are large, nevertheless it constitutes a 3.3 standard deviation from the background-only hypothesis.
The measured total cross section is also consistent with the NLO QCD predictions of $0.137\,$pb$\,\pm 11\%$ by Ref.~\cite{Garzelli:2011is} (CTEQ pdfs) 
or with $0.148\,$pb$\,\pm 11\%$ from our calculation (MSTW pdfs). 
In spite of the low number of events and correspondingly high statistical error, it is instructive to use this measured cross section to place bounds on the $\ttbZ$ couplings. 
This constitutes the first {\it direct} constraints on these couplings.
We perform a log-likelihood ratio analysis, as described in Section \ref{sect:analysis}. 
Our null hypothesis is derived from the experimental cross section $\sigma_{\ttbZ}^{\mathrm{CMS}}=0.28$~pb, from which we predict 24 events in the tri-lepton channel 
from $5~\invfb$ of data if no acceptance cuts are applied. 
We are adopting this number as our reference point against which we compare the alternative hypothesis, 
which is a given point in $(\ConeV,\ConeA)$ parameter space, and for which we calculate a predicted number of events in a $5~\invfb$ dataset.
To account for uncertainties in the theoretical prediction we include a uniform distribution spanned by the theoretical error band, 
for which we choose $\Delta_\mathrm{theor.\,unc.} = 40\%$ at LO and $\Delta_\mathrm{theor.\,unc.} = 15\%$ at NLO QCD.
For experimental systematics we need to introduce a Gaussian distributed probability. 
Hence, the function $\mathcal{G}(..|..)$ in Eq.~(\ref{errorfunctG}) has to be modified and becomes
\be
  \label{Gaussuncert}
  \mathcal{G} \left( \nu | \tilde{\nu},\tilde{\sigma} \right) = \frac{1}{\sqrt{2\pi\tilde{\sigma}^2}} \; e^{-(\nu-\tilde{\nu})^2 / (2\tilde{\sigma}^2)}, 
\ee
where $\tilde{\nu}$ is the mean value of the experimental measurement and $\tilde{\sigma}=\sigma_\mathrm{exp.\,syst.}$ is the systematic experimental uncertainty. 
In this way the mean value $\nu$ in the likelihood function is normal distributed during the generation of pseudo experiments with 
a standard deviation  $\sigma_\mathrm{exp.\,syst.}$. 
We adopt the experimental systematic error of $\sigma_\mathrm{exp.\,syst.}=\pm 20\%$ quoted in Ref.~\cite{Chatrchyan:2013qca}. 
Again, we use Eq.~(\ref{couplfit}) to generate a large grid corresponding to $\ConeV \in [-3.42:3.42]$ and $\ConeA \in [-3.61:3.61]$. 
Recall from Section~\ref{sect:ttzcoupl} that the SM couplings are $\ConeVSM \simeq 0.244$ and $\ConeASM \simeq -0.601$.

\begin{figure}[t]
\centering 
\includegraphics[width=0.44\textwidth]{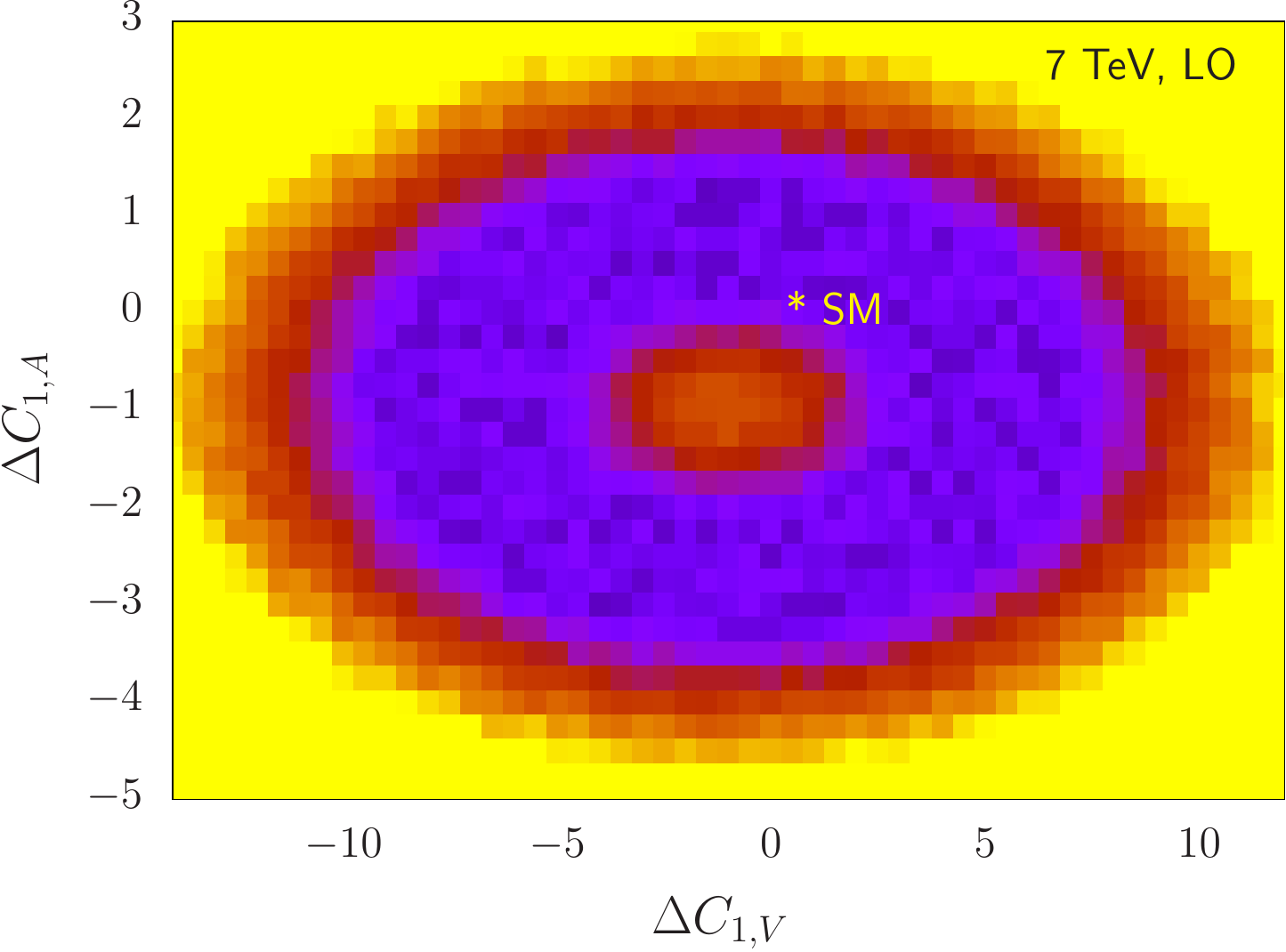}
\hspace{2mm}
\includegraphics[width=0.518\textwidth]{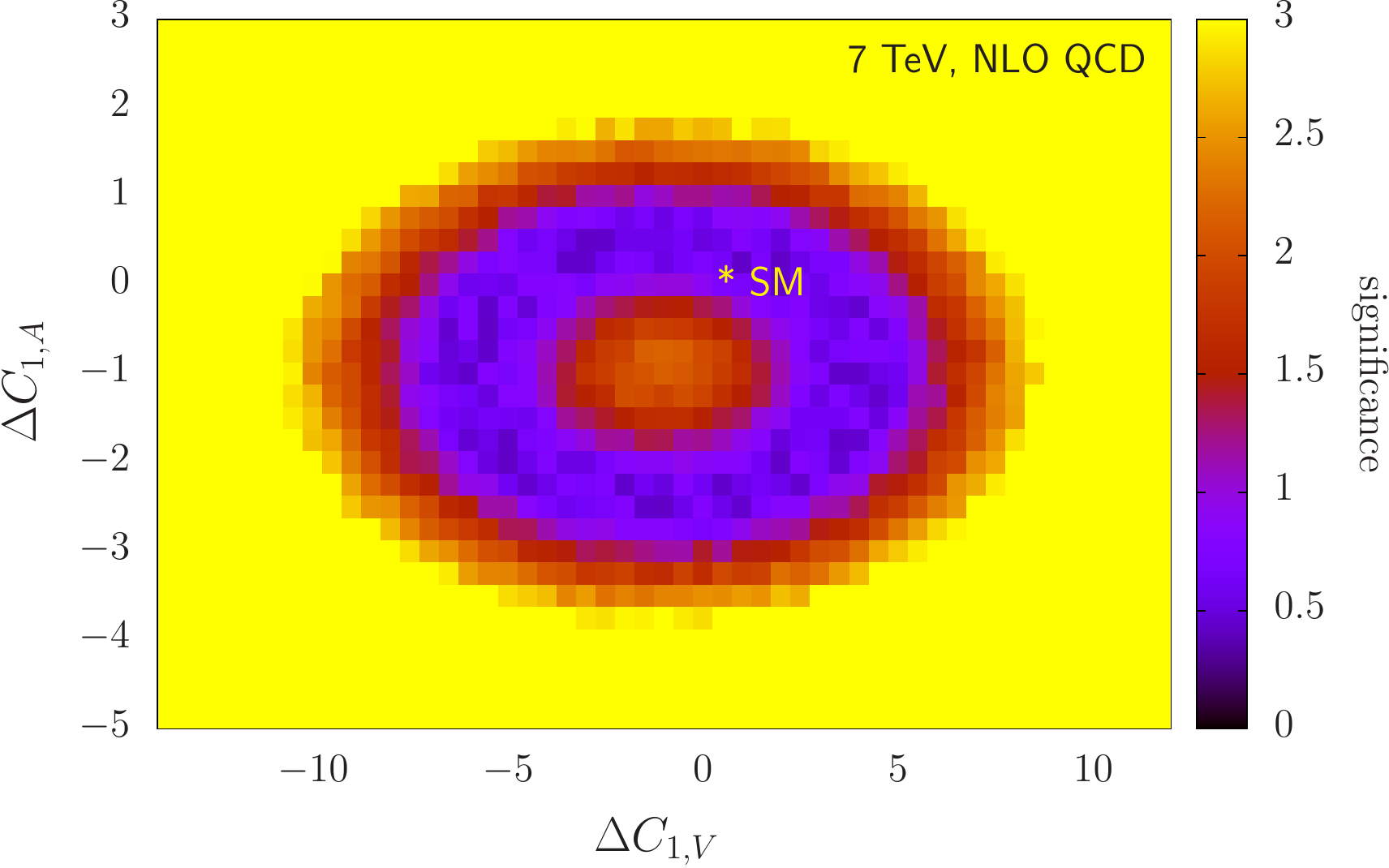}
\caption{\label{fig:CMSbounds}
Significance as a function of relative deviations for vector and axial couplings with respect to the SM value. 
The limits are obtained from the first measurement of the $\ttbZ$ cross section by CMS~\cite{Chatrchyan:2013qca}. 
The left~(right) plot shows the limits obtained from LO~(NLO QCD) input.
}
\end{figure}
The results of the log-likelihood ratio test are shown in Figure~\ref{fig:CMSbounds}, for our LO (left) and NLO (right) calculations.
The color code shows the significance with which an alternative coupling hypothesis can be excluded with respect to to the experimental data.
In the plane of relative deviations of vector and axial couplings, the point $(\DConeV,\DConeA)=(0.0,0.0)$ corresponds to the SM value.
Clearly, this point is fully consistent with the experimental measurement.
By comparing left and right plots we notice the stronger constraints when NLO input is used. 
The constraints from the data using a LO calculation are $ -11 \lesssim \DConeV \lesssim 10$ and $-4 \lesssim \DConeA \lesssim 2$ at the 95\% C.L.,
while they improve at NLO to $-8 \lesssim \DConeV \lesssim 7$ and $-3 \lesssim \DConeA \lesssim  1$.
Of course, these limits are extremely loose, and furthermore should be interpreted with care since very few events have been observed by the experiments so far.
Only a larger data set and detailed analysis of backgrounds and detector effects will provide more reliable constraints on the $\ttbZ$ couplings.
We nevertheless believe that these results are interesting to consider, especially when put into context with limits
obtained from the future high-energy LHC.

\begin{figure}[t]
\includegraphics[scale=0.503]{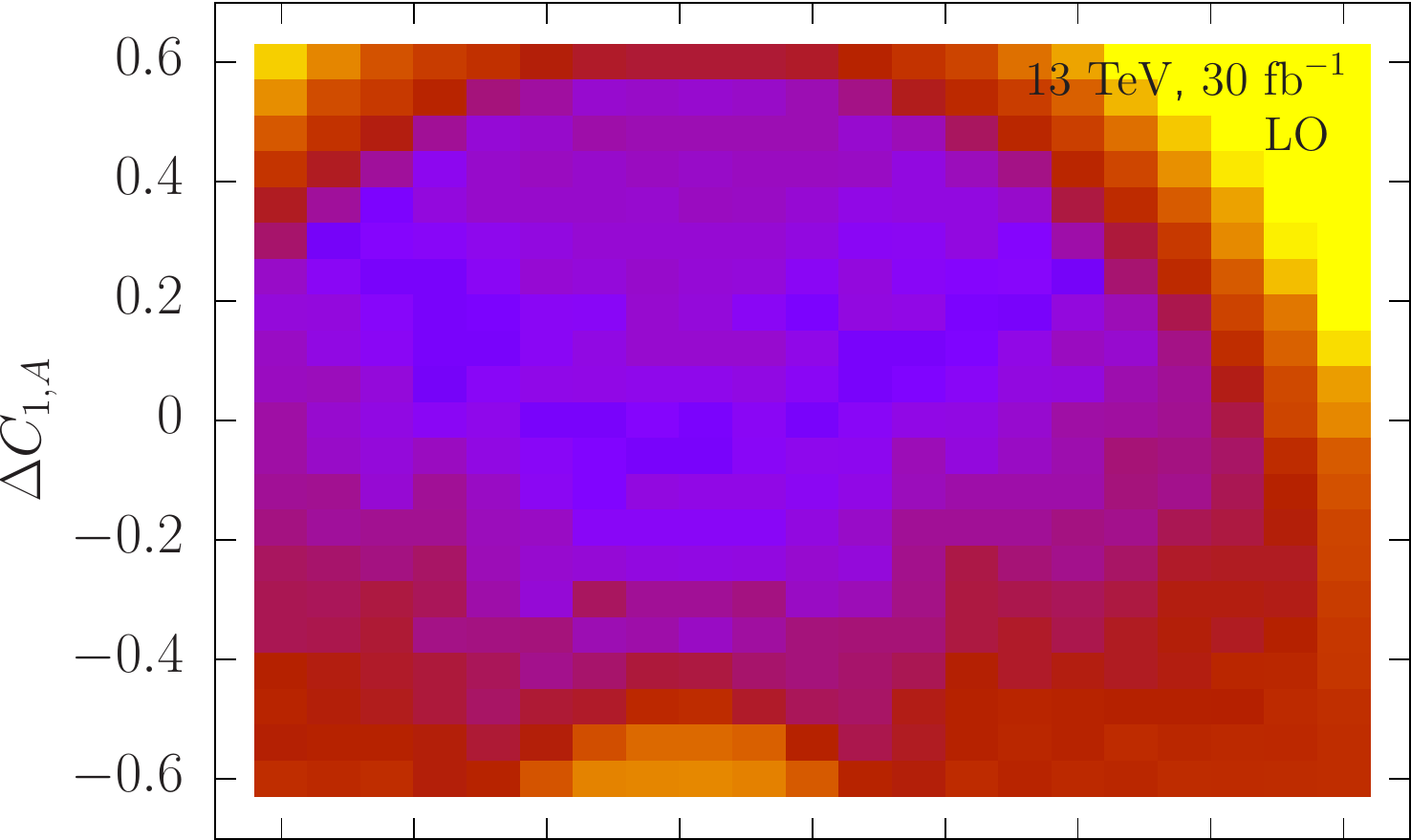} 
\includegraphics[scale=0.5]{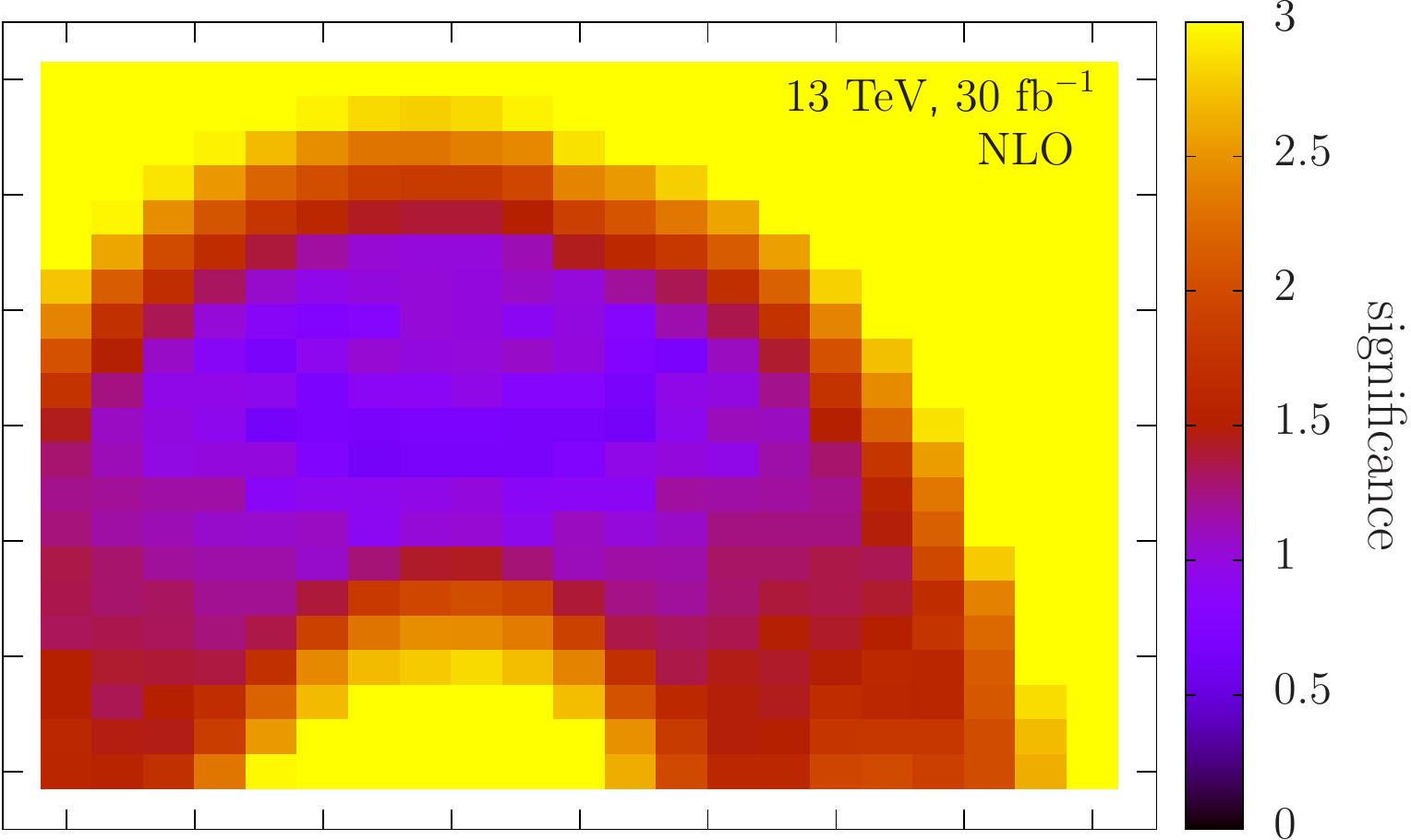} 
\\
\includegraphics[scale=0.503]{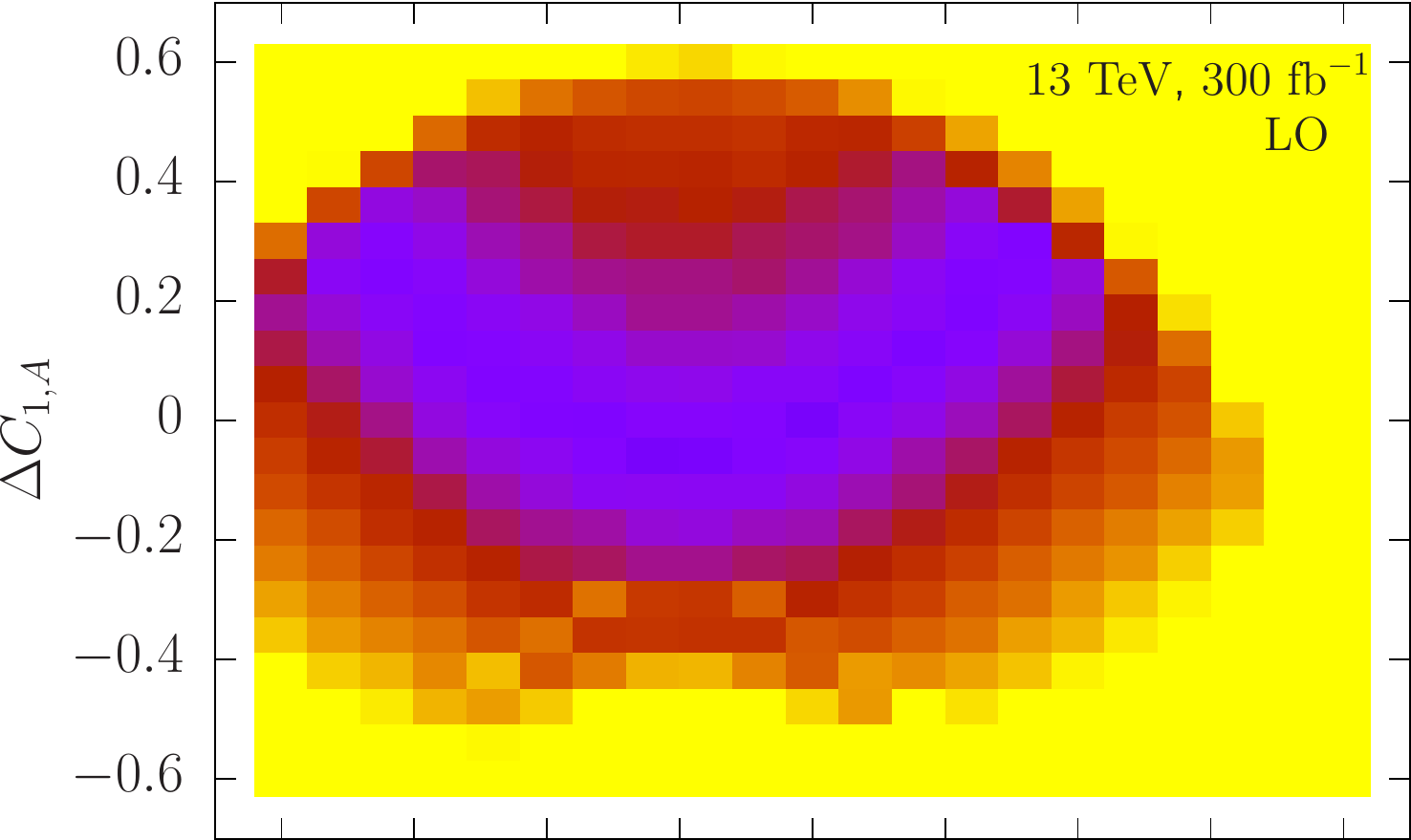} 
\includegraphics[scale=0.5]{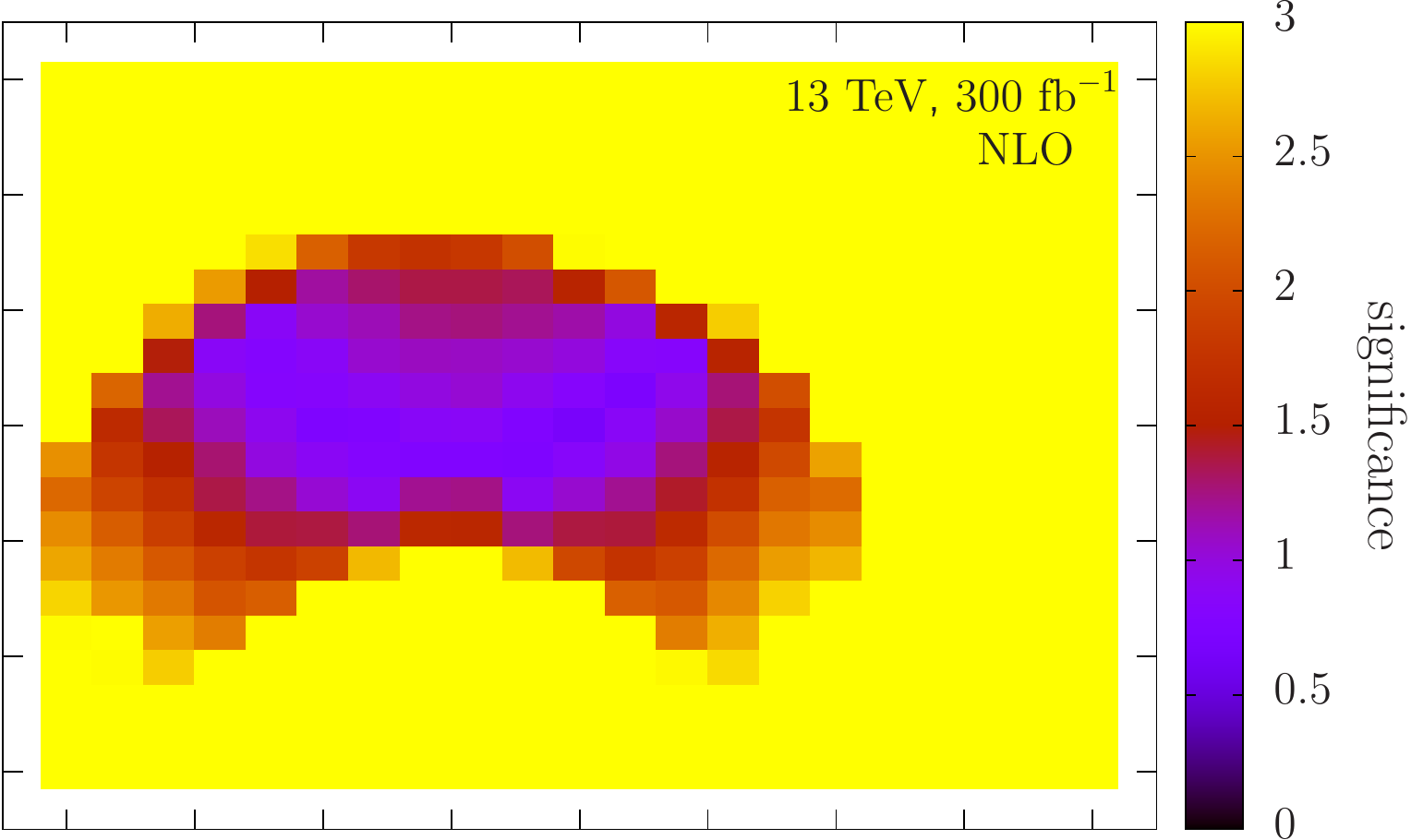} 
\\
\includegraphics[scale=0.503]{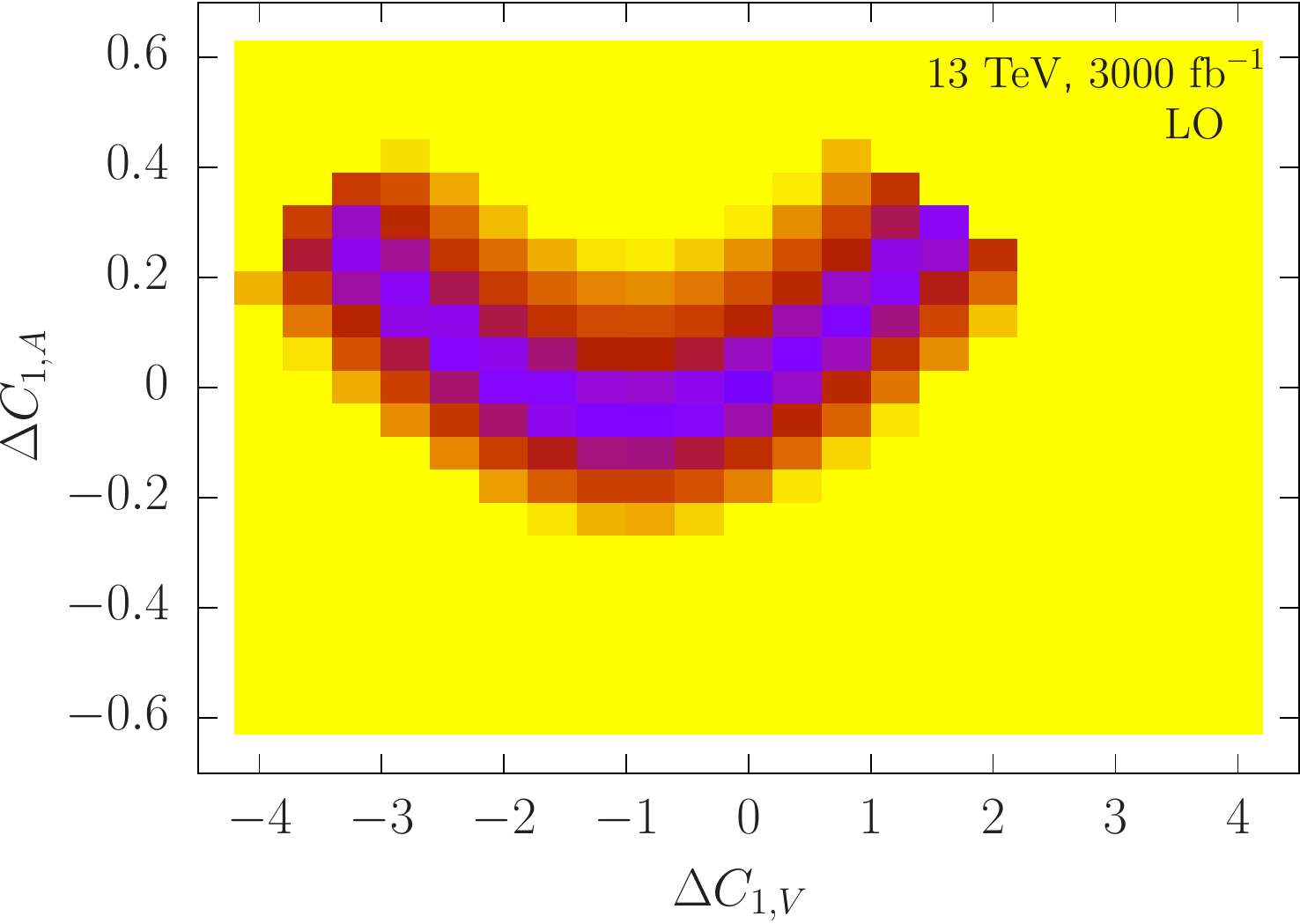} 
\includegraphics[scale=0.5]{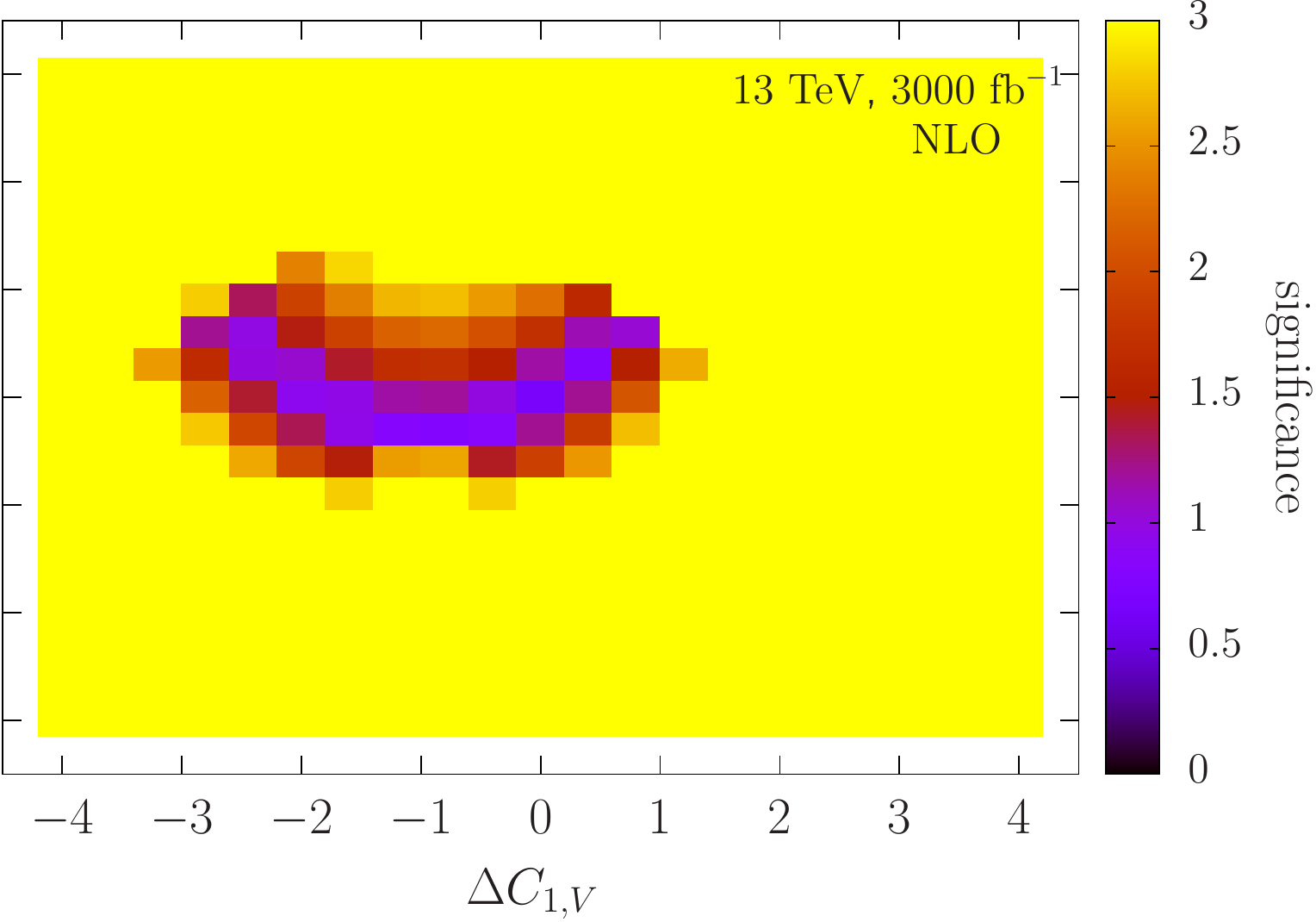} 
\caption{\label{fig:ix} Significance of deviations from the SM vector and axial couplings $\DConeV$ and $\DConeA$,  using 30, 300 and 3000 $\invfb$ of data at the $\sqrt{s}=13$ TeV LHC. 
Results using the LO prediction and uncertainty are shown on the left, the corresponding NLO QCD results are shown on the right hand side.}
\end{figure}

We turn now to this anticipated run of the LHC at $\sqrt{s}=13$~TeV.
Using the interpolation of Eq.~(\ref{couplfit}) we generate $441$ distributions in $\Delta \phi_{\ell\ell}$ for a grid of $21 \times 21$ $\DConeV,\DConeA$ couplings 
choices in the range $\pm 4$ and $\pm 0.6$, respectively. 
In terms of absolute numbers, this corresponds to varying $\ConeV \in [-0.732:1.22]$ and $\ConeA \in [-0.962:-0.240]$.
The plots in Fig.~\ref{fig:ix} show the significance with which non-SM $\ttbZ$ couplings can be separated from the SM hypothesis, assuming that the SM hypothesis is true.
Clearly, this significance is a function of the accumulated luminosity and the associated uncertainties at the given order in perturbation theory.
We therefore present six scenarios for luminosities of $30\, \invfb$, $300\, \invfb$, and $3000\, \invfb$ at the 13~TeV LHC with 
theory input at leading and next-to-leading order in QCD.
The LO pdf uncertainties are slightly smaller than at $\sqrt{s}=7$ TeV, allowing us to use an overall scale uncertainty of $30\%$ at LO and $15\%$ at NLO.
The couplings outside the light-blue area in Fig.~\ref{fig:ix} roughly correspond to the ones that can be excluded at 68\% confidence level (C.L.),
whereas couplings outside the orange colored boundary can be excluded at 95\% C.L.
From comparing the first, second and third row of plots in Fig.~\ref{fig:ix} it is immediately apparent that 
increasing the luminosity drastically improves the limits. 
By comparing plots in the left versus the right column we also see that 
the bounds at NLO QCD are far stronger.
This is a result of the reduced scale uncertainty and the larger cross section due to a positive perturbative correction.
Numerically, one finds that with $300\, \invfb$ and LO input, $\DConeV$ is constrained between $-4.0 < \DConeV < 2.8$ and $\DConeA$ is constrained between $-0.36 < \DConeA < 0.54$, 
at the 95\% C.L.\footnote{We checked that these LO limits roughly agree with the ones quoted in Ref.~\cite{Baur:2004uw} for the 14~TeV LHC.}
The limits improve with NLO QCD predictions to $-3.6 < \DConeV < 1.6$  and $-0.24 < \DConeA < 0.30$.
In terms of absolute values, these intervals correspond to $C_\mathrm{V}=0.24^{+0.39}_{-0.85}$ and $C_\mathrm{A}=-0.60^{+0.14}_{-0.18}$ at NLO QCD.
This is a reduction by 25\% and 42\%, respectively, compared to results obtained at leading order.
A noticeable feature in the exclusion limits at NLO, Fig.~\ref{fig:ix} (right), is the turnover from a 
downwards bend shape to an upwards bend shape when going from $30\,\invfb$ to $3000\,\invfb$.
This feature is a complicated effect of our uncertainty treatment and a transition from normalization to shape sensitivity in the exclusion.
In the upper right plot the exclusion region roughly follows the shape already observed in Fig.~\ref{fig:viii}. 
This can be understood from the fact that with a small event sample the exclusion limit is dominated by 
normalization differences, whereas different shapes have vanishing exclusion power. 
Using a larger event sample, shape sensitivity increases and allows us to exclude regions where normalization differences 
are small but shapes differ significantly.
This is the case in the lower right plot of Fig.~\ref{fig:ix}, where the previous downwards bend is safely excluded.

\subsection{Limits on dimension-six operators}
\label{DimSixLimits}

Having presented our main results in Fig.~\ref{fig:ix}, we can use the obtained limits to put constraints on 
possible effects of physics beyond the SM. 
The relevant dimension-six operators have been presented in Sect.~\ref{sect:ttzcoupl}.
This is also where we pointed out that the excellent agreement between experiment and prediction
for the $Z b_\mathrm{L} \bar{b}_\mathrm{L}$ couplings can be used 
(together with $\mathrm{SU(2)_L}$ symmetry of the SM) to relate $C^{(3,33)}_{\phi q} \approx - C^{(1,33)}_{\phi q}$.
In the following we will make use of this fact and eliminate $C^{(1,33)}_{\phi q}$ from our analysis
\footnote{It should be noted however that models exist which give vanishing corrections to $Zb\bar{b}$ for finite $C^{(3,33)}_{\phi q} + C^{(1,33)}_{\phi q}$. 
One example is given in Ref.~\cite{delAguila:2000rc} with vector-like quarks. In such case, our limits remain valid upon the 
replacement $C^{(3,33)}_{\phi q} \to C^{(3,33)}_{\phi q} -C^{(1,33)}_{\phi q} $.}.
Hence we are left with only two dimension-six operators, $C^{(3,33)}_{\phi q}$ and $C^{33}_{\phi u}$.
We begin by using the total cross section measurement of CMS at 7~TeV (see Sect.~\ref{sect:CouplLimits}). 
The limits on the total $\ttbZ$ cross section as a function of $\DConeV$ and $\DConeA$ directly
translate into limits on the two operators. 
Diagonalizing the dependence in Eq.~(\ref{Cone_NP}), we find at next-to-leading order
\be
  -1.35
  \le & \quad \frac{v^2}{\Lambda^2} \;  \mathrm{Re} \left[ C^{(3,33)}_{\phi q} \right]  \quad 
  &\le 
  0.68,
  \nonumber \\
  -2.34
  \le & \quad \frac{v^2}{\Lambda^2} \;  \mathrm{Re} \left[ C^{33}_{\phi u} \right]  \quad 
  &\le
  1.77. 
\ee
\begin{figure}[t]
\centering
\includegraphics[scale=0.43]{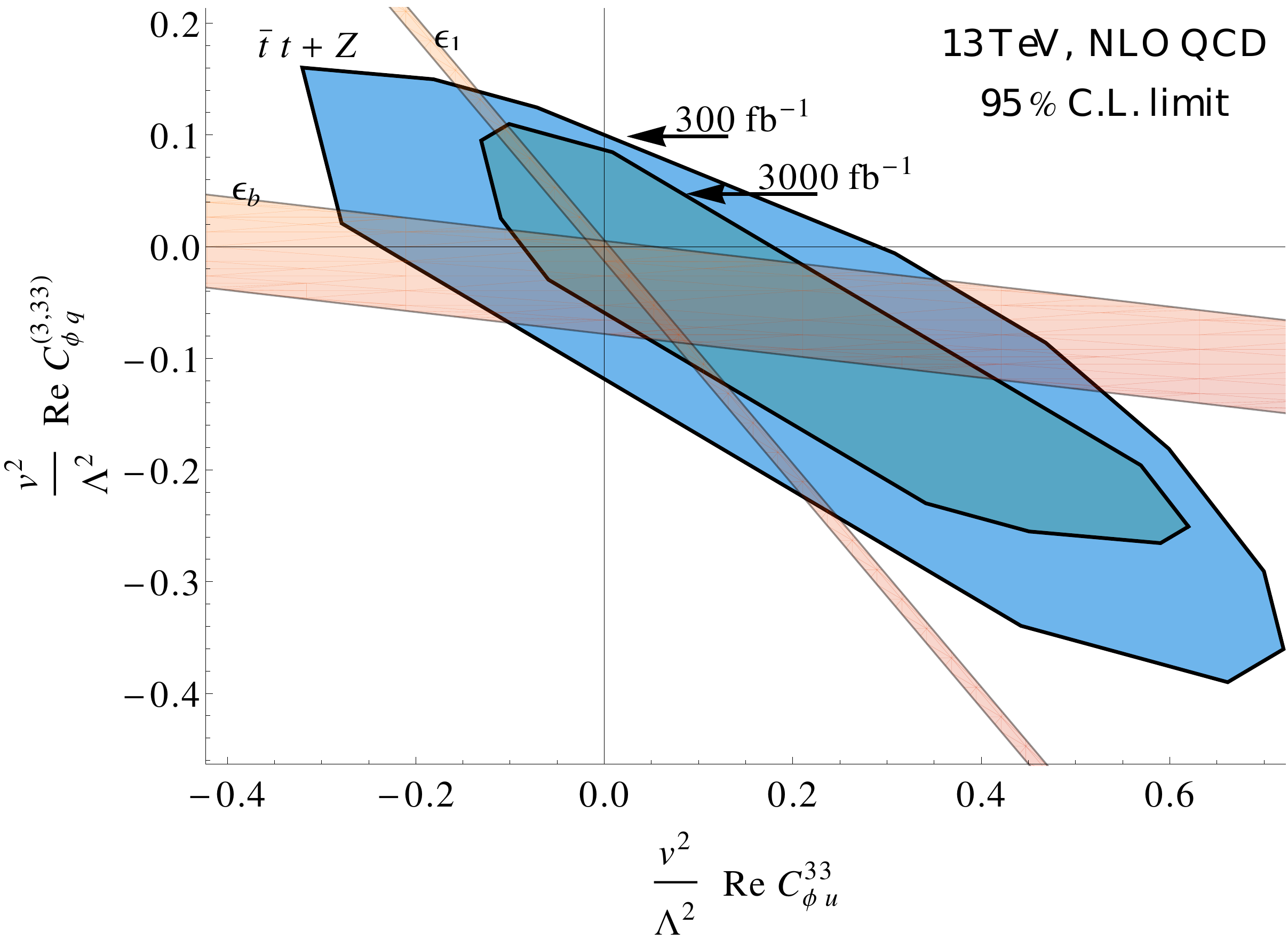} 
\caption{\label{fig:x} Projected constraints on the operators $C^{(33,3)}_{\phi q}$ and $C^{33}_{\phi u}$
obtained from the $\Delta \phi_{\ell^+_z \ell^-_z}$ distribution in $\ttbZ$ production at the 13~TeV LHC. 
The parameter space outside the blue colored area can be excluded at the 95\% C.L.
The thin bands are indirect constraints from electroweak precision data.
}
\end{figure}
This result should be considered with care given the low number of events observed in current experiments. 
More reliable and stringent limits can only be obtained once more data is accumulated.
To estimate how limits will improve in such a case, we use the results presented in Fig.~\ref{fig:ix}
for the luminosities 30, 300, and 3000~$\invfb$.
Recall that these results are not only based on the total cross section but also on the shapes
of the $\Delta \phi_{\ell^+_z \ell^-_z}$ distribution.
We find at leading order
\be
  \label{OPLimitsLO}
  \left. \begin{array}{l}
  -0.56 \\
  -0.44 \\
  -0.32  
  \end{array} \right\}   
  \le \quad \frac{v^2}{\Lambda^2} \;  \mathrm{Re} \left[ C^{(3,33)}_{\phi q} \right]  \quad 
  \le 
  \left\{ \begin{array}{ll}
  0.38 & \quad\quad \mathrm{with }~30  \, \invfb \\
  0.30 & \quad\quad \mathrm{with }~300 \, \invfb \\
  0.26 & \quad\quad \mathrm{with }~3000\, \invfb  
  \end{array} \right.
  ,\nonumber \\  \\ \nonumber
  \left. \begin{array}{l}
  -1.11 \\
  -0.54\\
  -0.29  
  \end{array} \right\} 
  \le \quad \frac{v^2}{\Lambda^2} \;  \mathrm{Re} \left[ C^{33}_{\phi u} \right]  \quad 
  \le 
  \left\{ \begin{array}{ll}
  1.02 & \quad\quad \mathrm{with }~30  \, \invfb \\
  0.93 & \quad\quad \mathrm{with }~300 \, \invfb \\
  0.88 & \quad\quad \mathrm{with }~3000\, \invfb  
  \end{array} \right. .
\ee
At next-to-leading order QCD the limits improve to
\be
  \label{OPLimitsNLO}
  \left. \begin{array}{l}
  -0.56 \\
  -0.40 \\
  -0.27  
  \end{array} \right\}   
  \le \quad \frac{v^2}{\Lambda^2} \;  \mathrm{Re} \left[ C^{(3,33)}_{\phi q} \right]  \quad 
  \le 
  \left\{ \begin{array}{ll}
  0.23 & \quad\quad \mathrm{with }~30  \, \invfb \\
  0.16 & \quad\quad \mathrm{with }~300 \, \invfb \\
  0.11 & \quad\quad \mathrm{with }~3000\, \invfb  
  \end{array} \right.
  ,\nonumber \\  \\ \nonumber
  \left. \begin{array}{l}
  -0.95 \\
  -0.35 \\
  -0.13  
  \end{array} \right\} 
  \le \quad \frac{v^2}{\Lambda^2} \;  \mathrm{Re} \left[ C^{33}_{\phi u} \right]  \quad 
  \le 
  \left\{ \begin{array}{ll}
  0.84 & \quad\quad \mathrm{with }~30  \, \invfb \\
  0.73 & \quad\quad \mathrm{with }~300 \, \invfb \\
  0.63 & \quad\quad \mathrm{with }~3000\, \invfb  
  \end{array} \right. .
\ee
Due to the weak correlation between vector and axial coupling limits observed in Fig.~\ref{fig:ix} and because of Eq.~(\ref{Cone_NP}),
the limits on $C^{(3,33)}_{\phi q}$ and $C^{33}_{\phi u}$ are strongly correlated. 
Hence, the results in Eqs.~(\ref{OPLimitsLO})-(\ref{OPLimitsNLO}) are very conservative.
A more appropriate graphical representation of these limits is given in Fig.~\ref{fig:x} for our NLO results. 
We also include the indirect constraints from electroweak precision observables $\varepsilon_1$ and $\varepsilon_b$~\cite{Altarelli:1990zd,Altarelli:1991fk,Altarelli:1993sz},
updated to account for $M_H=125.6$~GeV in Ref.~\cite{Ciuchini:2013pca}.
All effective operators outside the colored ellipse in Fig.~\ref{fig:x} can be excluded at the 95\% confidence level.
We observe that the limits from $\ttbZ$ production at the LHC are well-aligned with the precision limit from $\varepsilon_1$.
This can be understood from the fact that $\varepsilon_1$ is directly proportional to the SM $\rho$-parameter which receives sensitivity 
from the $Z$ boson self energy with a top quark loop insertion.
The constraint from $\varepsilon_b$ on the other hand arises from the measurement of $Z \to b \bar{b}$ and $\mathrm{SU(2)_L}$ symmetry of the SM.
Hence it leaves $C^{33}_{\phi u}$ mostly unconstrained since this operator contributes to the right handed current only.
Altogether, electroweak precision observables put very strong constraints on the $\ttbZ$ coupling;
however, these only arise through indirect sensitivity.
Only the analysis of $pp \to \ttbZ$ at the LHC will allow the placing of {\it direct} limits for the first time.
\\

As a final comment, we note that studies of the $\ttbZ$ couplings are also possible at a future lepton collider,
which provides a much cleaner environment and is therefore ideally suited to high-precision measurements.
In $e^+e^-$ collisions, sensitivity arises through the exchange of intermediate weak gauge bosons, $e^+e^- \to Z/\gamma^* \to \ttb$.
The production of a real $Z$ boson in association with $\ttb$ pairs is therefore not required.
This allows couplings to be studied close to the $\ttb$ threshold, as well as at higher energies, e.g. at $\sqrt{s}=500$~GeV \cite{Abe:2001swa,Adelman:2013gis,Amjad:2013tlv}.
The only disadvantage in $e^+e^-$ collisions is the presence of both $Z$ and $\gamma$ intermediate states. 
To disentangle their couplings to top quarks, experiments with variable beam polarization or different center-of-mass energies are required.
Prospects for $\ttbZ$ coupling limits are given in Ref.~\cite{Devetak:2010n} and, similarly, limits on higher dimensional operators 
as presented in this Section can be found in Ref.~\cite{AguilarSaavedra:2012vh}.
Generally, it is expected that the constraints improve by one order of magnitude, and possibly as much as two orders of the magnitude for the vector $\ttbZ$ coupling, 
compared to limits from the LHC. 

\section{Conclusion}
In this article we studied top quark pair production in association with a $Z$~boson.
Due to its relatively high production threshold and 
penalties from small branching fractions, 
this process was never observed at the Tevatron.
Even at the 7 and 8~TeV run of the LHC only a few candidate events were collected.
As a consequence there is no {\it direct} measurement of the top quark to $Z$~boson couplings to this date. 
This situation will change once the high energy LHC delivers its first tens~$\invfb$ of data.
We therefore study the process $pp\to\ttbZ$ in the tri-lepton final state,   
which provides the best compromise between clean signature and large enough cross section. 
The central question that we try to answer is by how much limits on
$\ttbZ$ couplings improve once NLO QCD predictions are used.
A particularly sensitive observable for such a study is the opening angle between the two leptons from the $Z$~boson decay.
We perform the analysis with a binned log-likelihood ratio test which proves advantageous for several reasons.
Firstly, the use of likelihood functions guarantees reliable results even for low number of events when, for example, a simple $\chi^2$ analysis would fail.
Secondly, non-Gaussian systematic errors such as theoretical scale uncertainties can be straightforwardly implemented in the likelihood ratio test.
In addition it turns out to be relatively easy to implement this procedure at NLO QCD.
We begin by performing this analysis on the inclusive cross section reported by CMS, which allows us to place the first direct constraints on the $\ttbZ$ couplings at the LHC. 
We proceed to an analysis at the $\sqrt{s}=13$ TeV LHC run.
Assuming a residual theoretical uncertainty of 15\% at NLO we find that with 300~$\invfb$ of data the vector and axial couplings can 
be constrained to $C_\mathrm{V}=0.24^{+0.39}_{-0.85}$ and $C_\mathrm{A}=-0.60^{+0.14}_{-0.18}$ at the 95\%~C.L.
This is a significant improvement compared to an analysis at leading order.
Even a first determination with only 30~$\invfb$ of data might be possible if NLO input is used, yielding 
limits which are about two times weaker. 
We also translate our constraints on vector and axial couplings into limits on dimension-six operators contributing to the $\ttbZ$ couplings beyond the SM.
The viable region for these operators can be significantly reduced with measurements of $pp \to \ttbZ$ and $\mathcal{O}(100)\invfb$ of data.
This allows us to contrast high precision indirect limits from electroweak observables with a direct determination from the LHC.

Finally, we note that effects of New Physics can modify the $\ttbZ$ coupling beyond vector and axial currents through $q^2$-dependent  higher dimensional operators. 
Those couplings typically introduce non-renormalizable interactions and require an extension of our one-loop integrand reduction method. 
This is an interesting subject for a continuation of this work.
Another interesting future topic is the study of sensitivity at a 100~TeV $pp$ collider or at an $e^+ e^-$ machine.
At any rate, we look forward to the first analysis of the $\ttb+Z,W,\gamma,H$ processes and the subsequent precision studies of top quark phenomenology.

\acknowledgments
We acknowledge helpful conversations with P.~Agrawal, J.~Campbell, R.K.~Ellis, Y.~Gao and N.~Tran. Fermilab is operated by Fermi Research Alliance, LLC under Contract No. De-AC02-07CH11359 with the United States Department of Energy. This research used resources of the National Energy Research Scientific Computing Center, which is supported by the Office of Science of the U.S. Department of Energy under Contract No. DE-AC02-05CH11231.

\bibliographystyle{JHEP}
\bibliography{ttbZ}

\providecommand{\href}[2]{#2}\begingroup\raggedright\begin{thebibliography}{10}

\bibitem{Chatrchyan:2012ufa}
{\bf CMS} Collaboration, S.~Chatrchyan {\em et.~al.}, {\it {Observation of a
  new boson at a mass of 125 GeV with the CMS experiment at the LHC}},  {\em
  Phys.Lett.} {\bf B716} (2012) 30--61
  [\href{http://arXiv.org/abs/1207.7235}{{\tt 1207.7235}}].

\bibitem{Aad:2012tfa}
{\bf ATLAS} Collaboration, G.~Aad {\em et.~al.}, {\it {Observation of a new
  particle in the search for the Standard Model Higgs boson with the ATLAS
  detector at the LHC}},  {\em Phys.Lett.} {\bf B716} (2012) 1--29
  [\href{http://arXiv.org/abs/1207.7214}{{\tt 1207.7214}}].

\bibitem{ATLAS:2011nka}
{\bf ATLAS} Collaboration, {\it {Measurement of the inclusive t tbar gamma
  cross section with the ATLAS detector}},  tech. rep., 2011.

\bibitem{CMS:2014wma}
{\bf CMS} Collaboration, {\it {Measurement of the inclusive top-quark pair +
  photon production cross section in the muon + jets channel in pp collisions
  at 8 TeV}},  tech. rep., 2014.

\bibitem{ATLAS-CONF-2012-126}
{\bf ATLAS} Collaboration, {\it {Search for $t\bar{t}Z$ production in the three
  lepton final state with $4.7$ ${\rm fb}^{-1}$ of $\sqrt{s}=7$ TeV $pp$
  collision data collected by the ATLAS detector}},  Tech. Rep.
  ATLAS-CONF-2012-126, CERN, Geneva, Aug, 2012.

\bibitem{Chatrchyan:2013qca}
{\bf CMS} Collaboration, S.~Chatrchyan {\em et.~al.}, {\it {Measurement of
  associated production of vector bosons and top quark-antiquark pairs at
  sqrt(s) = 7 TeV}},  {\em Phys.Rev.Lett.} {\bf 110} (2013) 172002
  [\href{http://arXiv.org/abs/1303.3239}{{\tt 1303.3239}}].

\bibitem{ALEPH:2005ab}
{\bf ALEPH, DELPHI, L3 , OPAL, SLD, LEP Electroweak Working Group, SLD
  Electroweak Group, SLD Heavy Flavour Group} Collaboration, S.~Schael {\em
  et.~al.}, {\it {Precision electroweak measurements on the $Z$ resonance}},
  {\em Phys.Rept.} {\bf 427} (2006) 257--454
  [\href{http://arXiv.org/abs/hep-ex/0509008}{{\tt hep-ex/0509008}}].

\bibitem{Abdallah:2008ab}
{\bf DELPHI} Collaboration, J.~Abdallah {\em et.~al.}, {\it {A Study of b
  anti-b Production in e+e- Collisions at s**(1/2) = 130-GeV - 207-GeV}},  {\em
  Eur.Phys.J.} {\bf C60} (2009) 1--15
  [\href{http://arXiv.org/abs/0901.4461}{{\tt 0901.4461}}].

\bibitem{Hollik:1988ii}
W.~Hollik, {\it {Radiative Corrections in the Standard Model and their Role for
  Precision Tests of the Electroweak Theory}},  {\em Fortsch.Phys.} {\bf 38}
  (1990) 165--260.

\bibitem{PhysRevD.82.055001}
T.~Ibrahim and P.~Nath, {\it Top quark electric dipole moment in a minimal
  supersymmetric standard model extension with vectorlike multiplets},  {\em
  Phys. Rev. D} {\bf 82} (Sep, 2010) 055001.

\bibitem{PhysRevD.84.015003}
T.~Ibrahim and P.~Nath, {\it Chromoelectric dipole moment of the top quark in
  models with vectorlike multiplets},  {\em Phys. Rev. D} {\bf 84} (Jul, 2011)
  015003.

\bibitem{Schmaltz:2002wx}
M.~Schmaltz, {\it {Physics beyond the standard model (theory): Introducing the
  little Higgs}},  {\em Nucl.Phys.Proc.Suppl.} {\bf 117} (2003) 40--49
  [\href{http://arXiv.org/abs/hep-ph/0210415}{{\tt hep-ph/0210415}}].

\bibitem{Cheng:2003ju}
H.-C. Cheng and I.~Low, {\it {TeV symmetry and the little hierarchy problem}},
  {\em JHEP} {\bf 0309} (2003) 051
  [\href{http://arXiv.org/abs/hep-ph/0308199}{{\tt hep-ph/0308199}}].

\bibitem{Frampton:1999xi}
P.~H. Frampton, P.~Hung and M.~Sher, {\it {Quarks and leptons beyond the third
  generation}},  {\em Phys.Rept.} {\bf 330} (2000) 263
  [\href{http://arXiv.org/abs/hep-ph/9903387}{{\tt hep-ph/9903387}}].

\bibitem{Dobrescu:2009vz}
B.~A. Dobrescu, K.~Kong and R.~Mahbubani, {\it {Prospects for top-prime quark
  discovery at the Tevatron}},  {\em JHEP} {\bf 0906} (2009) 001
  [\href{http://arXiv.org/abs/0902.0792}{{\tt 0902.0792}}].

\bibitem{Aguilar-Saavedra:2013qpa}
J.~Aguilar-Saavedra, R.~Benbrik, S.~Heinemeyer and M.~Perez-Victoria, {\it {A
  handbook of vector-like quarks: mixing and single production}},  {\em
  Phys.Rev.} {\bf D88} (2013) 094010
  [\href{http://arXiv.org/abs/1306.0572}{{\tt 1306.0572}}].

\bibitem{PhysRevD.86.095017}
R.~S. Chivukula, P.~Ittisamai, E.~H. Simmons, B.~Coleppa, H.~E. Logan,
  A.~Martin and J.~Ren, {\it Discovering strong top dynamics at the lhc},  {\em
  Phys. Rev. D} {\bf 86} (Nov, 2012) 095017.

\bibitem{Grojean:2013qca}
C.~Grojean, O.~Matsedonskyi and G.~Panico, {\it {Light top partners and
  precision physics}},  {\em JHEP} {\bf 1310} (2013) 160
  [\href{http://arXiv.org/abs/1306.4655}{{\tt 1306.4655}}].

\bibitem{Randall:1999ee}
L.~Randall and R.~Sundrum, {\it {A Large mass hierarchy from a small extra
  dimension}},  {\em Phys.Rev.Lett.} {\bf 83} (1999) 3370--3373
  [\href{http://arXiv.org/abs/hep-ph/9905221}{{\tt hep-ph/9905221}}].

\bibitem{Richard:2013pwa}
F.~Richard, {\it {Can LHC observe an anomaly in $t\bar tZ$ production?}},
  \href{http://arXiv.org/abs/1304.3594}{{\tt 1304.3594}}.

\bibitem{Baur:2004uw}
U.~Baur, A.~Juste, L.~Orr and D.~Rainwater, {\it {Probing electroweak top quark
  couplings at hadron colliders}},  {\em Phys.Rev.} {\bf D71} (2005) 054013
  [\href{http://arXiv.org/abs/hep-ph/0412021}{{\tt hep-ph/0412021}}].

\bibitem{Baur:2005wi}
U.~Baur, A.~Juste, D.~Rainwater and L.~Orr, {\it {Improved measurement of $ttZ$
  couplings at the CERN LHC}},  {\em Phys.Rev.} {\bf D73} (2006) 034016
  [\href{http://arXiv.org/abs/hep-ph/0512262}{{\tt hep-ph/0512262}}].

\bibitem{Berger:2009hi}
E.~L. Berger, Q.-H. Cao and I.~Low, {\it {Model Independent Constraints Among
  the Wtb, Zb anti-b, and Zt anti-t Couplings}},  {\em Phys.Rev.} {\bf D80}
  (2009) 074020 [\href{http://arXiv.org/abs/0907.2191}{{\tt 0907.2191}}].

\bibitem{Lazopoulos:2008de}
A.~Lazopoulos, T.~McElmurry, K.~Melnikov and F.~Petriello, {\it
  {Next-to-leading order QCD corrections to $t \bar{t} Z$ production at the
  LHC}},  {\em Phys.Lett.} {\bf B666} (2008) 62--65
  [\href{http://arXiv.org/abs/0804.2220}{{\tt 0804.2220}}].

\bibitem{Kardos:2011na}
A.~Kardos, Z.~Trocsanyi and C.~Papadopoulos, {\it {Top quark pair production in
  association with a Z-boson at NLO accuracy}},  {\em Phys.Rev.} {\bf D85}
  (2012) 054015 [\href{http://arXiv.org/abs/1111.0610}{{\tt 1111.0610}}].

\bibitem{Garzelli:2011is}
M.~Garzelli, A.~Kardos, C.~Papadopoulos and Z.~Trocsanyi, {\it {Z0 - boson
  production in association with a top anti-top pair at NLO accuracy with
  parton shower effects}},  {\em Phys.Rev.} {\bf D85} (2012) 074022
  [\href{http://arXiv.org/abs/1111.1444}{{\tt 1111.1444}}].

\bibitem{Garzelli:2012bn}
M.~Garzelli, A.~Kardos, C.~Papadopoulos and Z.~Trocsanyi, {\it {t $\bar{t}$
  $W^{+-}$ and t $\bar{t}$ Z Hadroproduction at NLO accuracy in QCD with Parton
  Shower and Hadronization effects}},  {\em JHEP} {\bf 1211} (2012) 056
  [\href{http://arXiv.org/abs/1208.2665}{{\tt 1208.2665}}].

\bibitem{Buttar:2008jx}
C.~Buttar, J.~D'Hondt, M.~Kramer, G.~Salam, M.~Wobisch {\em et.~al.}, {\it
  {Standard Model Handles and Candles Working Group: Tools and Jets Summary
  Report}},  \href{http://arXiv.org/abs/0803.0678}{{\tt 0803.0678}}.

\bibitem{Denner:2012yc}
A.~Denner, S.~Dittmaier, S.~Kallweit and S.~Pozzorini, {\it {NLO QCD
  corrections to off-shell top-antitop production with leptonic decays at
  hadron colliders}},  {\em JHEP} {\bf 1210} (2012) 110
  [\href{http://arXiv.org/abs/1207.5018}{{\tt 1207.5018}}].

\bibitem{Bevilacqua:2010qb}
G.~Bevilacqua, M.~Czakon, A.~van Hameren, C.~G. Papadopoulos and M.~Worek, {\it
  {Complete off-shell effects in top quark pair hadroproduction with leptonic
  decay at next-to-leading order}},  {\em JHEP} {\bf 1102} (2011) 083
  [\href{http://arXiv.org/abs/1012.4230}{{\tt 1012.4230}}].

\bibitem{Heinrich:2013qaa}
G.~Heinrich, A.~Maier, R.~Nisius, J.~Schlenk and J.~Winter, {\it {NLO QCD
  corrections to WWbb production with leptonic decays in the light of top quark
  mass and asymmetry measurements}},
  \href{http://arXiv.org/abs/1312.6659}{{\tt 1312.6659}}.

\bibitem{Campbell:2013yla}
J.~Campbell, R.~K. Ellis and R.~Rontsch, {\it {Single top production in
  association with a Z boson at the LHC}},  {\em Phys.Rev.} {\bf D87} (2013)
  114006 [\href{http://arXiv.org/abs/1302.3856}{{\tt 1302.3856}}].

\bibitem{Altarelli:2000nt}
G.~Altarelli, L.~Conti and V.~Lubicz, {\it {The t $\to$ WZ b decay in the
  standard model: A Critical reanalysis}},  {\em Phys.Lett.} {\bf B502} (2001)
  125--132 [\href{http://arXiv.org/abs/hep-ph/0010090}{{\tt hep-ph/0010090}}].

\bibitem{Decker:1992wz}
R.~Decker, M.~Nowakowski and A.~Pilaftsis, {\it {Dominant three-body decays of
  a heavy Higgs and top quark}},  {\em Z.Phys.} {\bf C57} (1993) 339--348
  [\href{http://arXiv.org/abs/hep-ph/9301283}{{\tt hep-ph/9301283}}].

\bibitem{Mahlon:1994us}
G.~Mahlon and S.~J. Parke, {\it {Finite width effects in top quark decays}},
  {\em Phys.Lett.} {\bf B347} (1995) 394--398
  [\href{http://arXiv.org/abs/hep-ph/9412250}{{\tt hep-ph/9412250}}].

\bibitem{Jenkins:1996zd}
E.~E. Jenkins, {\it {The Rare top decays $t \to b W^{+} Z$ and $t \to c W^{+}
  W^{-}$}},  {\em Phys.Rev.} {\bf D56} (1997) 458--466
  [\href{http://arXiv.org/abs/hep-ph/9612211}{{\tt hep-ph/9612211}}].

\bibitem{Ossola:2006}
G.~Ossola, C.~G. Papadopoulos and R.~Pittau, {\it {Reducing full one-loop
  amplitudes to scalar integrals at the integrand level}},  {\em Nucl.Phys.}
  {\bf B763} (2007) 147--169 [\href{http://arXiv.org/abs/hep-ph/0609007}{{\tt
  hep-ph/0609007}}].

\bibitem{Ellis:2007br}
R.~K. Ellis, W.~Giele and Z.~Kunszt, {\it {A Numerical Unitarity Formalism for
  Evaluating One-Loop Amplitudes}},  {\em JHEP} {\bf 0803} (2008) 003
  [\href{http://arXiv.org/abs/0708.2398}{{\tt 0708.2398}}].

\bibitem{Giele:2008ve}
W.~T. Giele, Z.~Kunszt and K.~Melnikov, {\it {Full one-loop amplitudes from
  tree amplitudes}},  {\em JHEP} {\bf 0804} (2008) 049
  [\href{http://arXiv.org/abs/0801.2237}{{\tt 0801.2237}}].

\bibitem{Ellis:2008ir}
R.~K. Ellis, W.~T. Giele, Z.~Kunszt and K.~Melnikov, {\it {Masses, fermions and
  generalized $D$-dimensional unitarity}},  {\em Nucl.Phys.} {\bf B822} (2009)
  270--282 [\href{http://arXiv.org/abs/0806.3467}{{\tt 0806.3467}}].

\bibitem{Ellis:2011}
R.~K. Ellis, Z.~Kunszt, K.~Melnikov and G.~Zanderighi, {\it {One-loop
  calculations in quantum field theory: from Feynman diagrams to unitarity
  cuts}},  {\em Phys.Rept.} {\bf 518} (2012) 141--250
  [\href{http://arXiv.org/abs/1105.4319}{{\tt 1105.4319}}].

\bibitem{Melnikov:2009dn}
K.~Melnikov and M.~Schulze, {\it {NLO QCD corrections to top quark pair
  production and decay at hadron colliders}},  {\em JHEP} {\bf 0908} (2009) 049
  [\href{http://arXiv.org/abs/0907.3090}{{\tt 0907.3090}}].

\bibitem{Catani:1996vz}
S.~Catani and M.~Seymour, {\it {A General algorithm for calculating jet
  cross-sections in NLO QCD}},  {\em Nucl.Phys.} {\bf B485} (1997) 291--419
  [\href{http://arXiv.org/abs/hep-ph/9605323}{{\tt hep-ph/9605323}}].

\bibitem{Catani:2002hc}
S.~Catani, S.~Dittmaier, M.~H. Seymour and Z.~Trocsanyi, {\it {The Dipole
  formalism for next-to-leading order QCD calculations with massive partons}},
  {\em Nucl.Phys.} {\bf B627} (2002) 189--265
  [\href{http://arXiv.org/abs/hep-ph/0201036}{{\tt hep-ph/0201036}}].

\bibitem{Nagy:1998bb}
Z.~Nagy and Z.~Trocsanyi, {\it {Next-to-leading order calculation of four jet
  observables in electron positron annihilation}},  {\em Phys.Rev.} {\bf D59}
  (1999) 014020 [\href{http://arXiv.org/abs/hep-ph/9806317}{{\tt
  hep-ph/9806317}}].

\bibitem{Nagy:2003tz}
Z.~Nagy, {\it {Next-to-leading order calculation of three jet observables in
  hadron hadron collision}},  {\em Phys.Rev.} {\bf D68} (2003) 094002
  [\href{http://arXiv.org/abs/hep-ph/0307268}{{\tt hep-ph/0307268}}].

\bibitem{Bevilacqua:2009zn}
G.~Bevilacqua, M.~Czakon, C.~Papadopoulos, R.~Pittau and M.~Worek, {\it
  {Assault on the NLO Wishlist: pp $\to$ t anti-t b anti-b}},  {\em JHEP} {\bf
  0909} (2009) 109 [\href{http://arXiv.org/abs/0907.4723}{{\tt 0907.4723}}].

\bibitem{Campbell:2010ff}
J.~M. Campbell and R.~Ellis, {\it {MCFM for the Tevatron and the LHC}},  {\em
  Nucl.Phys.Proc.Suppl.} {\bf 205-206} (2010) 10--15
  [\href{http://arXiv.org/abs/1007.3492}{{\tt 1007.3492}}].

\bibitem{Melnikov:2011ta}
K.~Melnikov, M.~Schulze and A.~Scharf, {\it {QCD corrections to top quark pair
  production in association with a photon at hadron colliders}},  {\em
  Phys.Rev.} {\bf D83} (2011) 074013
  [\href{http://arXiv.org/abs/1102.1967}{{\tt 1102.1967}}].

\bibitem{pvegas}
R.~Kreckel, {\it {Parallel version of G.P.Lepage's VEGAS-algorithm}},  {\em
  Unpublished} (2000).

\bibitem{mpi-2-standard}
{Message Passing Interface Forum}, {\it {MPI2}: A message passing interface
  standard},  {\em High Performance Computing Applications} {\bf 12} (1998),
  no.~1--2 1--299.

\bibitem{Stelzer:1994ta}
T.~Stelzer and W.~Long, {\it {Automatic generation of tree level helicity
  amplitudes}},  {\em Comput.Phys.Commun.} {\bf 81} (1994) 357--371
  [\href{http://arXiv.org/abs/hep-ph/9401258}{{\tt hep-ph/9401258}}].

\bibitem{Cullen:2011ac}
G.~Cullen, N.~Greiner, G.~Heinrich, G.~Luisoni, P.~Mastrolia {\em et.~al.},
  {\it {Automated One-Loop Calculations with GoSam}},  {\em Eur.Phys.J.} {\bf
  C72} (2012) 1889 [\href{http://arXiv.org/abs/1111.2034}{{\tt 1111.2034}}].

\bibitem{Pumplin:2002vw}
J.~Pumplin, D.~Stump, J.~Huston, H.~Lai, P.~M. Nadolsky {\em et.~al.}, {\it
  {New generation of parton distributions with uncertainties from global QCD
  analysis}},  {\em JHEP} {\bf 0207} (2002) 012
  [\href{http://arXiv.org/abs/hep-ph/0201195}{{\tt hep-ph/0201195}}].

\bibitem{Nadolsky:2008zw}
P.~M. Nadolsky, H.-L. Lai, Q.-H. Cao, J.~Huston, J.~Pumplin {\em et.~al.}, {\it
  {Implications of CTEQ global analysis for collider observables}},  {\em
  Phys.Rev.} {\bf D78} (2008) 013004
  [\href{http://arXiv.org/abs/0802.0007}{{\tt 0802.0007}}].

\bibitem{AlcarazMaestre:2012vp}
J.~Alcaraz~Maestre {\em et.~al.}, {\it {The SM and NLO Multileg and SM MC
  Working Groups: Summary Report}},  \href{http://arXiv.org/abs/1203.6803}{{\tt
  1203.6803}}.

\bibitem{AguilarSaavedra:2008zc}
J.~Aguilar-Saavedra, {\it {A Minimal set of top anomalous couplings}},  {\em
  Nucl.Phys.} {\bf B812} (2009) 181--204
  [\href{http://arXiv.org/abs/0811.3842}{{\tt 0811.3842}}].

\bibitem{Zhang:2012cd}
C.~Zhang, N.~Greiner and S.~Willenbrock, {\it {Constraints on Non-standard Top
  Quark Couplings}},  {\em Phys.Rev.} {\bf D86} (2012) 014024
  [\href{http://arXiv.org/abs/1201.6670}{{\tt 1201.6670}}].

\bibitem{Bernabeu:1995gs}
J.~Bernabeu, D.~Comelli, L.~Lavoura and J.~P. Silva, {\it {Weak magnetic dipole
  moments in two Higgs doublet models}},  {\em Phys.Rev.} {\bf D53} (1996)
  5222--5232 [\href{http://arXiv.org/abs/hep-ph/9509416}{{\tt
  hep-ph/9509416}}].

\bibitem{Hollik:1998vz}
W.~Hollik, J.~I. Illana, S.~Rigolin, C.~Schappacher and D.~Stockinger, {\it
  {Top dipole form-factors and loop induced CP violation in supersymmetry}},
  {\em Nucl.Phys.} {\bf B551} (1999) 3--40
  [\href{http://arXiv.org/abs/hep-ph/9812298}{{\tt hep-ph/9812298}}].

\bibitem{Mastrolia:2012bu}
P.~Mastrolia, E.~Mirabella and T.~Peraro, {\it {Integrand reduction of one-loop
  scattering amplitudes through Laurent series expansion}},  {\em JHEP} {\bf
  1206} (2012) 095 [\href{http://arXiv.org/abs/1203.0291}{{\tt 1203.0291}}].

\bibitem{Ciuchini:2013pca}
M.~Ciuchini, E.~Franco, S.~Mishima and L.~Silvestrini, {\it {Electroweak
  Precision Observables, New Physics and the Nature of a 126 GeV Higgs Boson}},
   {\em JHEP} {\bf 1308} (2013) 106 [\href{http://arXiv.org/abs/1306.4644}{{\tt
  1306.4644}}].

\bibitem{Altarelli:1990zd}
G.~Altarelli and R.~Barbieri, {\it {Vacuum polarization effects of new physics
  on electroweak processes}},  {\em Phys.Lett.} {\bf B253} (1991) 161--167.

\bibitem{Altarelli:1991fk}
G.~Altarelli, R.~Barbieri and S.~Jadach, {\it {Toward a model independent
  analysis of electroweak data}},  {\em Nucl.Phys.} {\bf B369} (1992) 3--32.

\bibitem{Altarelli:1993sz}
G.~Altarelli, R.~Barbieri and F.~Caravaglios, {\it {Nonstandard analysis of
  electroweak precision data}},  {\em Nucl.Phys.} {\bf B405} (1993) 3--23.

\bibitem{Larios:1999au}
F.~Larios, M.~Perez and C.~Yuan, {\it {Analysis of $t b W$ and $t t Z$
  couplings from CLEO and LEP / SLC data}},  {\em Phys.Lett.} {\bf B457} (1999)
  334--340 [\href{http://arXiv.org/abs/hep-ph/9903394}{{\tt hep-ph/9903394}}].

\bibitem{Martin:2009iq}
A.~Martin, W.~Stirling, R.~Thorne and G.~Watt, {\it {Parton distributions for
  the LHC}},  {\em Eur.Phys.J.} {\bf C63} (2009) 189--285
  [\href{http://arXiv.org/abs/0901.0002}{{\tt 0901.0002}}].

\bibitem{Cacciari:2008gp}
M.~Cacciari, G.~P. Salam and G.~Soyez, {\it {The Anti-k(t) jet clustering
  algorithm}},  {\em JHEP} {\bf 0804} (2008) 063
  [\href{http://arXiv.org/abs/0802.1189}{{\tt 0802.1189}}].

\bibitem{Lai:2010vv}
H.-L. Lai, M.~Guzzi, J.~Huston, Z.~Li, P.~M. Nadolsky {\em et.~al.}, {\it {New
  parton distributions for collider physics}},  {\em Phys.Rev.} {\bf D82}
  (2010) 074024 [\href{http://arXiv.org/abs/1007.2241}{{\tt 1007.2241}}].

\bibitem{Cowan:2010js}
G.~Cowan, K.~Cranmer, E.~Gross and O.~Vitells, {\it {Asymptotic formulae for
  likelihood-based tests of new physics}},  {\em Eur.Phys.J.} {\bf C71} (2011)
  1554 [\href{http://arXiv.org/abs/1007.1727}{{\tt 1007.1727}}].

\bibitem{Feldman:1997qc}
G.~J. Feldman and R.~D. Cousins, {\it {A Unified approach to the classical
  statistical analysis of small signals}},  {\em Phys.Rev.} {\bf D57} (1998)
  3873--3889 [\href{http://arXiv.org/abs/physics/9711021}{{\tt
  physics/9711021}}].

\bibitem{NeymanPearsonLemma}
J.~Neyman and E.~S. Pearson, {\it {On the Problem of the Most Efficient Tests
  of Statistical Hypotheses}},  {\em Philosophical Transactions of the Royal
  Society} {\bf 231} (1933) 694--706.

\bibitem{Conway:2011in}
J.~Conway, {\it {Incorporating Nuisance Parameters in Likelihoods for
  Multisource Spectra}},  \href{http://arXiv.org/abs/1103.0354}{{\tt
  1103.0354}}.

\bibitem{Kuhn:2011ri}
J.~H. Kuhn and G.~Rodrigo, {\it {Charge asymmetries of top quarks at hadron
  colliders revisited}},  {\em JHEP} {\bf 1201} (2012) 063
  [\href{http://arXiv.org/abs/1109.6830}{{\tt 1109.6830}}].

\bibitem{Hollik:2011ps}
W.~Hollik and D.~Pagani, {\it {The electroweak contribution to the top quark
  forward-backward asymmetry at the Tevatron}},  {\em Phys.Rev.} {\bf D84}
  (2011) 093003 [\href{http://arXiv.org/abs/1107.2606}{{\tt 1107.2606}}].

\bibitem{delAguila:2000rc}
F.~del Aguila, M.~Perez-Victoria and J.~Santiago, {\it {Observable
  contributions of new exotic quarks to quark mixing}},  {\em JHEP} {\bf 0009}
  (2000) 011 [\href{http://arXiv.org/abs/hep-ph/0007316}{{\tt
  hep-ph/0007316}}].

\bibitem{Abe:2001swa}
{\bf American Linear Collider Working Group} Collaboration, T.~Abe {\em
  et.~al.}, {\it {Linear Collider Physics Resource Book for Snowmass 2001 -
  Part 3: Studies of Exotic and Standard Model Physics}},
  \href{http://arXiv.org/abs/hep-ex/0106057}{{\tt hep-ex/0106057}}.

\bibitem{Adelman:2013gis}
J.~Adelman, B.~Alvarez~Gonzalez, Y.~Bai, M.~Baumgart, R.~K. Ellis {\em
  et.~al.}, {\it {Top Couplings: pre-Snowmass Energy Frontier 2013 Overview}},
  \href{http://arXiv.org/abs/1309.1947}{{\tt 1309.1947}}.

\bibitem{Amjad:2013tlv}
M.~Amjad, M.~Boronat, T.~Frisson, I.~Garcia, R.~Poschl {\em et.~al.}, {\it {A
  precise determination of top quark electro-weak couplings at the ILC
  operating at $\sqrt{s}=500$ GeV}},
  \href{http://arXiv.org/abs/1307.8102}{{\tt 1307.8102}}.

\bibitem{Devetak:2010n}
E.~Devetak, A.~Nomerotski and M.~Peskin, {\it {Top quark anomalous couplings at
  the International Linear Collider}},  {\em Phys.Rev.} {\bf D84} (2011) 034029
  [\href{http://arXiv.org/abs/1005.1756}{{\tt 1005.1756}}].

\bibitem{AguilarSaavedra:2012vh}
J.~Aguilar-Saavedra, M.~Fiolhais and A.~Onofre, {\it {Top Effective Operators
  at the ILC}},  {\em JHEP} {\bf 1207} (2012) 180
  [\href{http://arXiv.org/abs/1206.1033}{{\tt 1206.1033}}].

\end{thebibliography}\endgroup

\end{document}